\documentclass[aps,prb,twocolumn,floatfix,showpacs,
superscriptaddress]{revtex4}
\usepackage{amsmath}
\usepackage[dvips]{graphicx}
\usepackage{subfigure}
\usepackage[colorlinks=true,linkcolor=blue,citecolor=red]{hyperref}

\newcommand{\eq}[2]
{\begin{equation}
    #1
    \label{#2}
  \end{equation}}

\begin{document}
\title{
Asymmetry between the electron- and hole-doped Mott \\
transition in the periodic Anderson model}

\author{G. Sordi}
\affiliation{Laboratoire de Physique des Solides, CNRS-UMR8502,
Universit\'e de Paris-Sud, Orsay 91405, France.}
\author{A. Amaricci}
\affiliation{Laboratoire de Physique des Solides, CNRS-UMR8502,
Universit\'e de Paris-Sud, Orsay 91405, France.}
\author{M.J. Rozenberg}
\affiliation{Laboratoire de Physique des Solides, CNRS-UMR8502,
Universit\'e de Paris-Sud, Orsay 91405, France.}
\affiliation{Departamento de F\'{\i}sica, FCEN, Universidad de Buenos
Aires, Ciudad Universitaria Pab.I, Buenos Aires (1428), Argentina.}

\date{\today}
\begin{abstract}
We study the doping driven Mott metal-insulator transition (MIT) in the
periodic Anderson model set in the Mott-Hubbard regime. 
A striking asymmetry for electron or hole driven transitions is found. 
The electron doped MIT at larger U is similar to the one found 
in the single band Hubbard model, with a first order character 
due to coexistence of solutions. 
The hole doped MIT, in contrast, is second order 
and can be described as the delocalization of Zhang-Rice singlets.

\end{abstract}

\maketitle

\section{INTRODUCTION}\label{sec:one}

The nature of the Mott transition, 
i.e. the metal-insulator transition 
driven by electronic correlations, 
is a central problem in physics 
of strongly correlated electrons systems.
The relevance of the problem 
was initially emphasized by Mott \cite{mott} 
in the 40's, 
trying to explain why some materials 
with odd electrons per the unit cell, 
like NiO, are insulators. 
In a Mott metal-insulator transition (MIT), 
a metallic system 
with a partially filled electron band 
suddenly opens an insulating gap. 
In practice the transition is usually driven 
by temperature, applied pressure, or chemical doping. 
The origin of the mechanism 
is in the correlation effects 
due to the strong on-site Coulomb repulsion 
experienced by electrons 
occupying rather localized orbitals, 
such as $d$ in transition metal oxides 
or $f$ in heavy fermion compounds.

The classical example of an experimental system 
exhibiting a Mott transition 
is vanadium oxide V$_2$O$_3$, 
which has received continuous attention 
since the pioneering work of McWhan \cite{mcwhan3}
in the 70's. 
That compound has a finite temperature 
first order metal-insulator transition 
that terminates at a high temperature 
2nd order critical point, in analogy with 
the finite liquid-gas transition line in water. 

From a theoretical perspective, 
it is considered that the Hubbard model, 
which contains a tight binding band 
plus an interaction term 
that describes local Coulomb repulsion 
between electrons occupying the same site, 
is a minimal Hamiltonian that may capture 
the basic physics of the Mott MIT. 
The most significant work on this model 
was initiated by Hubbard \cite{hubbard64}
in the 60's, where, 
starting from the insulating state 
at large interaction values, 
he described how the system 
may close the correlation gap 
as the bandwidth is increased 
to values of the order of 
the Coulomb repulsion energy. 
Later, in the 70's Brinkman and Rice \cite{br}, 
using a variational approach, 
started from the metallic state 
and showed how 
it can be destroyed 
by increasing correlation effects when 
the interaction strength 
becomes of the order of the bandwidth. 
Finally in the 90's the theoretical development
of the dynamical mean-field theory 
(DMFT) \cite{metznervoll,rmp} 
allowed to get new insight on this problem.
In the scenario for the Mott transition 
realized in the DMFT solution of the Hubbard model, 
for low temperatures and moderate interaction, 
the half filled Mott insulator 
can be driven to a correlated metallic state 
through a {\sl first order} transition \cite{rmp}. 
This transition can occur as a function of 
correlation strength, 
temperature 
or doping.
The solution of the Hubbard model within DMFT 
provided not only a connection 
between the approaches of Hubbard and Brinkman-Rice 
by showing how the system evolves 
from a metal to an insulator, 
but also produced a detailed description 
of the basic experimental phenomenology 
observed in the V$_2$O$_3$ compound \cite{rmp}.
In addition, it was later shown that 
the MIT can be described in terms of 
a Ginzburg-Landau scenario \cite{rck,klr,gkepjb} 
with theoretical predictions for the critical behavior 
of observables near the 2nd order critical point, 
that were eventually confirmed by experiments \cite{limelette}.

The physics of the Mott transitions, 
especially those driven by doping at low temperatures, 
became of unparalleled interest 
in condensed matter physics with the discovery 
of the high temperature cuprate superconductors \cite{ift} 
in the 80's 
and, in smaller but also significant measure, 
by the discovery of the non-Fermi liquid behavior 
in heavy fermion systems \cite{stewart}. 
In those systems, 
the effect of strong correlations is undisputed, 
since the active electronic degrees of freedom involve 
the localized $d$ and $f$ orbitals. 
Therefore, these systems are identified 
as doped Mott insulators, 
however their phase diagrams 
and the evolution of their physical properties 
cannot be associated to the DMFT scenario 
for the Mott transition 
that was so successfully applied to V$_2$O$_3$. 

Besides the Hubbard model, 
the periodic Anderson model (PAM) 
is another minimal Hamiltonian that is often investigated 
in the context of strongly correlated electron systems. 
That model contains two orbitals per unit cell, 
one local with on-site Coulomb repulsion 
and the other non-interacting and itinerant. 
At each lattice site, the two orbitals are hybridized. 
This model is more realistic than the Hubbard, 
since it describes with greater detail 
the actual situation in real compounds. 
For instance, in transition metal oxides 
where the overlaps between neighboring oxygen $p-$orbitals 
provide itineracy to the electrons, 
while the localized $d-$orbitals of the transition metal 
experience the stronger correlation effects. 

Despite the higher degree of detail included in the PAM, 
it is often assumed in theoretical approaches 
that the physical behavior of the PAM 
would result qualitatively similar 
to that of the Hubbard model at low frequencies. 
That statement can be mathematically justified 
in certain parameter regimes, 
however its general validity is less evident 

The DMFT is a theoretical approach 
that is mathematically exact in the limit 
of large lattice dimensionality \cite{metznervoll,rmp} 
which has been extensively used to study 
the Mott transition in the Hubbard model, 
and, to a lesser degree, has also been employed
to investigate the physics of the PAM \cite{jarrell,mgk,marcelo95,jap}. 
Therefore, in the light of the previous discussion, 
a natural question to address 
is whether within DMFT 
the Mott transition scenario of the PAM 
is indeed qualitatively similar 
to that of the Hubbard model or, 
if contrary to usual expectations, 
it brings about new physical behaviors. 
This issue has been the focus 
of our recent investigations \cite{sar},
where we showed that in fact a different type 
of doping driven Mott MIT 
is realized in the PAM, even in a parameter regime 
whether it might be expected that the identification 
with the Hubbard model may hold. 
The present study 
extends and provides further details to that work. 
In particular, we present new data that illustrate the
different behavior of physical quantities on the two
sides of the transition, we provide new comparisons of the
Green functions to results using a $T=0$ numerical technique, and we
extend the discussion of the 
physical origin of two transitions. 
We should clarify that there is a qualitatively different insulator
state that can be realized in the PAM, namely, the Kondo insulator. It
is obtained in the particle-hole symmetric case, where the total
occupation is {\em even} (ie, 2 electrons per site). The Kondo
insulator is qualitatively different from a Mott insulator, because it
is due to band hybridization effect and correlations merely serve to
reduce the hybridization gap. This insulator has a temperature-driven
crossover to a metallic state that has been investigated within DMFT
using QMC techniques \cite{held,heldbulla}. In contrast, the Mott insulator
state that we consider here corresponds to a state with {\em odd}
total number of particles and its physical origin is entirely due to
strong correlation effects.

The paper is organized as follows. 
In Sec.~\ref{sec:methodology} we introduce the PAM 
and justify the choice of the parameter regime. 
We also summarize the DMFT equations 
and provide details on the numerical techniques 
we use to solve the associated impurity problem.  
In Sec.~\ref{sec:results} we present the results 
and discuss the Mott transitions found in the PAM. 
In Sec.~\ref{sec:discussion} we present a discussion 
of the physical origin of the different scenario for the MIT 
found in the PAM with respect to the HM. 
In Sec.\ref{sec:conclusions} we present the conclusions.

\section{METHODOLOGY}\label{sec:methodology}

\subsection{Model}\label{subsec:model}
\begin{figure}%[!ht]
\centering
\includegraphics[width=8cm,clip]{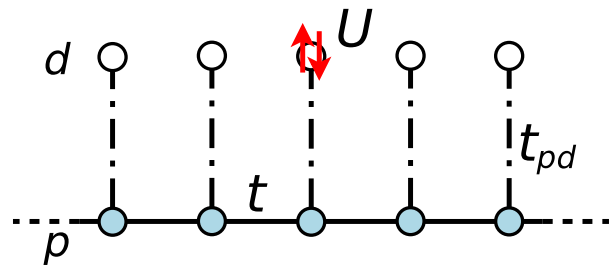}
\caption{Schematic representation 
of the periodic Anderson model 
for the case of a one dimensional chain. 
At each site of the lattice, a $d$ orbital (open circle) 
hybridizes with a $p$ orbital (full circle) 
through the amplitude $t_{pd}$ (dot-dashed line). 
Two electrons in the $d$ orbital experience 
a Coulomb repulsion $U$.
The hopping amplitude between the $p$ orbitals 
at neighboring sites $i$ and $j$ is $t_{ij}=t$ (solid line).
}
\label{fig1}
\end{figure}

The Hamiltonian of the PAM is given by
\eq{
\begin{split}
H=&-\sum_{<ij>\sigma} t_{ij} (p^+_{i\sigma}p_{j\sigma} + p^+_{j\sigma}p_{i\sigma}) +
\left(\epsilon_p - \mu \right) \sum_{i\sigma} p^+_{i\sigma}p_{i\sigma} \\
 &+(\epsilon_d-\mu)\sum_{i\sigma} d^+_{i\sigma}d_{i\sigma}
 +t_{pd} \sum_{i\sigma}\left( d^+_{i\sigma}p_{i\sigma} +p^+_{i\sigma}d_{i\sigma}
\right) \\
&+U\sum_i
\left(n_{di\uparrow}-\tfrac{1}{2}\right)\left(n_{di\downarrow}-\tfrac{1}{2}
\right).
\end{split}
}{PAMHam}
Here $p_{i\sigma}$ and $p_{i\sigma}^+$ operators 
destroy and create electrons at $p$ orbitals on site $i$ 
with spin $\sigma$. 
The $p$ orbitals have site energy $\epsilon_p$ 
and overlap via the hopping term $t_{ij}=t$ to form a band.
$d_{i\sigma}$ and $d_{i\sigma}^+$ operators 
destroy and create electrons at $d$ orbitals on site $i$ 
with spin $\sigma$.
The local $d$ orbitals have site energy $\epsilon_d$ 
and are hybridized to the $p$ orbitals 
with the constant on-site amplitude $t_{pd}$.
$U$ is the energy cost of double occupation of the $d$ orbitals 
at each site and $\mu$ is the chemical potential. 
$\Delta_0=\epsilon_d-\epsilon_p$, the difference between 
the $d$ level and the center of the $p$ band, defines the bare 
charge-transfer energy.
Fig.~\ref{fig1} shows a schematic representation 
of the Hamiltonian.

The insulator solutions of the PAM have been studied in literature mostly 
in the symmetric regime $\Delta_0=0$, $\mu=0$, 
where the system is a Kondo (or renormalized-band) insulator. 
In the present work, we shall focus on a different and less explored parameter regime, where the system is in a Mott insulator state.
A key difference between these insulator states is that the former is realized at an even total electron occupation ($n_{tot}=2$), while the latter is realized at an odd occupation ($n_{tot}=1$ or $n_{tot}=3$).
Yet, the Mott insulator state may still be classified as a charge-transfer insulator or Mott-Hubbard insulator, according to the Zaanen-Sawatzky-Allen scheme \cite{zsa}.

The action associated with the 
Hamiltonian (\ref{PAMHam}) reads
\eq{
\begin{split}
{\rm S}=&
- \sum_{k,\sigma} \int_0^{\beta}d\tau \int_0^\beta d\tau'  \psi^+_{k\sigma}(\tau) \hat{G}_{0\sigma}^{-1}(\tau - 
\tau')\psi_{k\sigma}(\tau') \\
&+U\sum_{i} \int_0^{\beta} d\tau
\left[n_{id\uparrow}(\tau)-\tfrac{1}{2}\right]\left[n_{id\downarrow}(\tau)
-\tfrac{1}{2}\right],
\end{split}
}{S_lattice}
where $\psi_{k\sigma}=\{d_{\sigma},p_{k\sigma}\}$, 
$\psi_{k\sigma}^+=\{d^+_{\sigma},p^+_{k\sigma}\}$ 
and the inverse matrix propagator 
$\hat{G}_0^{-1}$ is given by:
\begin{equation}
\hat{G}_{0\sigma}^{-1} (k,i\omega_n)= 
\left( \begin{array}{cc}
 i\omega_n+\mu-\epsilon_d & t_{pd} \\
t_{pd} & i\omega_n+\mu-\epsilon_p -\epsilon_k
           \end{array} \right),
\end{equation}
where $\epsilon_k$ is the Fourier transform 
of the hopping term $t$. 
The lattice Green's function $\hat{G}$ is then written 
using the Dyson equation $\hat{G}^{-1}=\hat{G}_0^{-1}-\hat{\Sigma}$,
where 
\begin{equation}
\hat{\Sigma}_{\sigma}(k,i\omega_n)=
\left( \begin{array}{cc}
\Sigma_{\sigma}(k,i\omega_n) & 0 \\
0 & 0
           \end{array} \right)
\end{equation}
and $\Sigma_{\sigma}(k,i\omega_n)$
is the $d-$electron self-energy.
Here we are interested in a magnetically disordered state, 
thus the spin index can be dropped. 
In this case, the lattice Green's functions 
for the $p$ and $d$ electrons are explicitly given by:
\eq{
\begin{split}
G_{pp}^{-1}(k, i\omega_n) = &  i\omega_n +\mu-\epsilon_p -\epsilon_k \\ 
 & -\frac{t_{pd}^2}{i\omega_n +\mu-\epsilon_d 
-\Sigma(k,i\omega_n)} 
\end{split}}{}
\eq{
\begin{split}
G_{dd}^{-1} (k, i\omega_n) = & i\omega_n +\mu-\epsilon_d  -\Sigma(k,i\omega_n) \\
 & -\frac{t_{pd}^2}{i\omega_n +\mu-\epsilon_p 
-\epsilon_k}.
\end{split}}{latticeGF_full}

The local Green's functions are then obtained 
performing the integration over momenta,
\begin{equation}
G_{\alpha}(i\omega_n) = \frac{1}{N} \sum_{k} G_{\alpha}(k, i\omega_n)
=\int \rho_0(\epsilon) G_{\alpha}(\epsilon,i\omega_n) 
d\epsilon,
\end{equation}
where $\alpha={pp,dd}$ 
and $\rho_{0}(\epsilon)=\sum_k \delta(\epsilon-\epsilon_k)$ 
is the free ($U=0$ and $t_{pd}=0$) density of states of the $p$ electrons.

The PAM has some simply solvable limits 
such as of vanishing hybridization, $t_{pd}=0$, 
or of vanishing correlation strength, $U=0$.
For the latter case, 
the PAM describes two hybridized one-particle bands, 
that are obtained diagonalizing the Hamiltonian and read:
\begin{align}
E_{\pm} (k)=\frac{1}{2}\left(\epsilon_d +\epsilon_p +\epsilon_k -2\mu \pm  
\sqrt{(\epsilon_k-\Delta_0)^2+4t_{pd}^2}\,\right).
\label{poles_u0}
\end{align}
\begin{figure}[ht]
\centering
\includegraphics[width=8cm,clip]{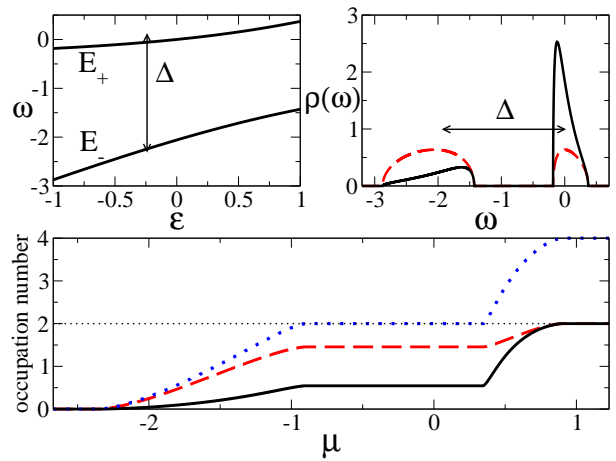}
\caption{
Top left panel: 
solid lines are the two branches $E_{\pm}(\epsilon)$ 
for $U=0$, $\Delta_0=1$, $t_{pd}=0.9$ 
and $\mu=0.529$, that gives $n_{tot}=3$. 
Top right panel: 
density of states for the $p$ and $d$ electrons 
(dashed and solid line), 
for the same model parameters as in the left panel.
Note that the effective distance between the bands, 
$\Delta\approx2$, results to be larger than 
the bare value $\Delta_0=\epsilon_d-\epsilon_p=1$.
The free ($U=0$ and $t_{pd}=0$) DOS 
of the conduction electrons is semi-elliptical 
with with a half-bandwidth equal to unity. 
Bottom panel: 
particle occupation 
$n_d$ (solid line), $n_p$ (dashed) and $n_{tot}$ (dotted) 
as a function of the chemical potential 
for $U=0$, $\Delta_0=1$, $t_{pd}=0.9$.}
\label{fig2}
\end{figure}

From the many interesting parameter regimes 
that this model has, 
we shall focus our study on a particular one 
where the low energy physics of the PAM 
is a priori expected to correspond 
to that of the Hubbard model. 
Thus, we consider the case where 
the localized $d$ orbital is near the Fermi energy 
and with an occupation close to one, 
while the $p$ orbital band is well beneath in energy 
and almost fully occupied. 
By virtue of the hybridization term, 
the $d$ electrons acquire a finite dispersion 
and form a narrow band that crosses the Fermi energy. 
It has a bandwidth $\sim t_{pd}^2/\Delta$, 
where $\Delta$ is the distance 
between the two hybridized bands, 
$\Delta \approx E_{+}-E_{-} > \Delta_{0}$, and
is subject to strong correlation effects 
when the on-site Coulomb term is turned on. 

For reference, the 
solution for the non-interacting case in the
chosen parameter regime is shown in Fig.~\ref{fig2}. 
We set the units adopting a model semicircular
density of states for the $p$-electrons 
with half-bandwidth $D=2t=1$ 
As we shall see in the next section, this density of
states is actually realized in a Bethe lattice in the
limit of infinite spatial dimensions.
In the upper panels we show 
the single particle dispersion $E_{\pm}$  
and the resulting density of states 
for $\Delta_0=1$, $t_{pd}=0.9$ and $\mu=0.529$, 
that gives a total occupation per site $n_{tot}$ 
equal to 3. 
In the lower panel we plot the total occupation 
$n_{tot}=n_p +n_d$ together with the partial occupation.
The plateaux in the curves 
signal incompressible states 
that correspond to insulators. 
These are observed at $n_{tot}=2$ and $n_{tot}=4$. 
In the first case, $n_{tot}$ equal to 2, 
one has a hybridization gap insulator, 
since the band $E_{-}$ is full and $E_{+}$ is empty. 
For $n_{tot}=4$ the state is of a full band insulator. 
For $2<n_{tot}<4$ the system is metallic, 
since the lower band with mostly $p$ character is full 
and the narrow band with mainly $d$ character is partially filled.

We shall now focus on the effects of correlation 
on such a metallic case, 
realized for $n_{tot}=3$,
where the lower electron band with mostly $p$ character 
is close to being full (i.e. is occupied by two electrons) 
and the narrow $d$ electron band is half-filled 
(i.e. with an occupation $n_d$ close to 1). 
In the chosen parameter regime, 
there is a single band crossing the Fermi energy 
which has mostly $d$ electron character. 
So correlations effects will affect it strongerly. 
For values of the interaction $U$ 
larger than its bandwidth $\sim t_{pd}^2/\Delta$ 
one may expect that a correlation gap would open 
and the system becomes a Mott insulator.
This is the regime where 
the identification of the low energy physics 
of the PAM with the one band Hubbard model may hold.

\subsection{DMFT and the limit of infinite dimensions}
\label{dmft_dinfty}

To go beyond this qualitative discussion 
we need to obtain reliable solutions of the model Hamiltonian 
in the strongly interacting regime. 
We thus recur to the DMFT formulation 
where exact numerical methods 
can be used to solve the problem \cite{rmp}.

The DMFT solution becomes exact 
in the limit of large spatial dimensionality \cite{metznervoll}
or, equivalently, large lattice connectivity $z$. 
For this limit to remain physical one is required 
to rescale the hopping $t$ amplitude as $t/\sqrt{z}$, 
so that the density of states 
$\rho_0(\epsilon)=\sum_k \delta(\epsilon-\epsilon_k)$ 
gives a finite value for the mean kinetic energy \cite{metznervoll}.
As is well known in DMFT, the specific lattice structure is not essential, 
and several lattice types could be used. 
For instance, 
the free (i.e., $t_{pd}=0$, $U=0$) density of states 
of the hypercubic lattice \cite{metznervoll,jap}, 
which is the generalization of the square lattice 
to the limit of high $z$, reads:
\begin{equation}
\rho_0^{\rm hyper}(\epsilon)= \frac{1}{t\sqrt{\pi}} \exp{\left(-\frac{\epsilon^2}{2t^2}\right)}, 
\label{dos_HCL}
\end{equation}
where $\epsilon$ denotes 
the noninteracting single particle energy.
Another lattice type which is often adopted 
is the Bethe lattice, 
whose density of states reads,
\begin{equation}
\rho_0(\epsilon)=\frac{1}{2\pi t^2} \sqrt{4t^2-\epsilon^2}.
\label{dos_BL}
\end{equation}
In the following we shall adopt 
this type of lattice structure, 
as it is better suited for some of the numerical methods 
that we shall employ.
As unit of energy  we set the half-bandwidth 
of the Bethe lattice semi-circular DOS, $D=2t=1$.
The key mathematical simplification 
arising from the $z\rightarrow\infty$ limit 
is the locality of the self-energy, 
i.e. its $k$ independence.
Thus, there is no longer need to keep the momentum label 
in the single particle energies of the band structure $\epsilon_k$, 
and the energy $\epsilon$ itself is simply kept 
as the quantum number.

\subsection{Mean-field equations}\label{dmft_equations}

In the limit of large lattice connectivity $z\rightarrow \infty$, 
the PAM can be exactly mapped 
onto a single impurity Anderson model 
supplemented with a self consistency condition.
The derivation of the DMFT equations 
has already been presented in detail 
elsewhere \cite{rmp,antoine,jeschke_pam}, 
so here we shall just briefly summarize the main steps 
and the final expressions. 

A direct way to derive the DMFT equations 
is to apply the cavity method \cite{rmp}. 
The key idea is to focus on a given (any) site of the lattice 
and to integrate out the degrees of freedom 
on all the other lattice sites 
in order to obtain the local effective action at the selected site. 
In doing that, one shall also obtain a self-consistency condition 
which restores the translational invariance that was (temporarily) 
broken with the selection of a given lattice site.
After integration of all sites other than the one selected, 
the local effective action is obtained,
\eq{
\begin{split}
{\rm S_{\rm eff}}=&
-\int_0^{\beta}d\tau \int_0^\beta d\tau' \sum_{\sigma} \psi^+_{\sigma}(\tau)
\hat{{\cal G}}^{-1}_0(\tau - \tau')\psi_{\sigma}(\tau') \\
&+U\int_0^{\beta} d\tau
\left[n_{d\uparrow}(\tau)-\tfrac{1}{2}\right]\left[n_{d\downarrow}(\tau)
-\tfrac{1}{2}\right],
\end{split}
}{Seff}
where $\psi_{\sigma}=\{d_\sigma,p_\sigma\}$, 
$\psi_{\sigma}^+=\{d_\sigma^+,p_\sigma^+\}$ 
correspond to the two atomic orbitals 
of the given (arbitrary) site of the lattice. 
The local inverse propagator reads
\eq{
\hat{\cal{G}}_0^{-1} (i\omega_n)= \left( \begin{array}{cc}
i\omega_n+\mu-\epsilon_d & t_{pd} \\
t_{pd} & i\omega_n+\mu-\epsilon_p -t^2 \tilde{G}_{pp}
           \end{array} \right),
}{cavity}
where $\tilde{G}_{pp}$ is the cavity Green's function 
that encodes the information of the propagation
of electrons in the lattice,
restricted not to return to the local site.
These two equations define 
the so called associated impurity problem of the model.

In order to restore the translational invariance 
and to obtain a closed set of equations, 
one has to relate the local inverse propagator ${\cal G}_0$ 
to the Green's function of the original lattice.  
In the Bethe lattice this relation is simple \cite{rmp}: 
the $p-$electron cavity Green's function $\tilde{G}_{pp}$ 
becomes the local $p-$electron Green's function, $\tilde{G}_{pp}=G_{pp}$.
From this relation and (\ref{cavity}), 
one obtains the self-consistency condition for 
the impurity problem. 
It can be casted only in terms of $[{\cal G}_0]_{dd}$ 
and reads,
\eq{
\begin{split}
[{\cal G}_0^{-1}]_{dd}(i\omega_n) = & 
i\omega_n+\mu-\epsilon_d \\
 & -\frac{{{t^2_{pd}}}}{i\omega_n+\mu-\epsilon_p-t^2G_{pp}}\ .
\end{split}}{self}

In practice an iterative procedure is implemented 
to solve the set of DMFT equations: 
given an ansatz for $[{\cal G}_0]_{dd}$, 
and the fact that the interactions are local 
and only act on the $d$ orbital,
the impurity many-body problem (\ref{Seff}) 
can be solved to produce 
a local $d-$electron Green's function 
$G_{dd}=-\langle dd^+ \rangle_{S_{\rm eff}}$.
This defines a local self-energy 
$\Sigma=[{\cal G}_0^{-1}]_{dd}-G_{dd}^{-1}$,
that allows for the calculation of 
the local $p-$electron Green's function $G_{pp}$ 
(\ref{latticeGF_full})
\eq{
G_{pp}(i\omega_n)=\int{ \frac{\rho_0(\epsilon) \,d\epsilon}{i\omega_n +\mu -\epsilon_p -
\frac{t^2_{pd}}{i\omega_n+\mu-\epsilon_d -\Sigma(i\omega_n)} - \epsilon}}.
}{gpp1}
The obtained $G_{pp}$ and $\Sigma$ are then used 
as input to the self-consistency condition Eq.~\ref{self} 
to produce a new $[{\cal G}_0]_{dd}$. 
This process is iterated until convergence is reached. 
{\it At the self-consistent point}, 
the Green's functions $G_{dd}$, $G_{pp}$ 
and the self-energy $\Sigma$
correspond to the local propagators of the original lattice model. 
Moreover, since in the $z \to \infty$ limit $\Sigma$ is local, 
then all the $k$ dependent (or $\epsilon$ dependent) 
propagators of the original lattice 
can also be computed from this self-energy.

It is useful rewrite the local Green's functions 
in terms of the Hilbert transform of the density of states 
$\tilde{D}(\xi)=\int_{-\infty}^{\infty}d\epsilon\rho_0(\epsilon)/(\xi-\epsilon)$. 
For the conduction electron Green's function 
$G_{pp}$ we get
\eq{
\begin{split}
G_{pp} (i\omega_n)=& \tilde{D} \left(i\omega_n +\mu -\epsilon_p -\frac{t_{pd}^2}{i\omega_n +\mu -\epsilon_d  -\Sigma 
(i\omega_n)}\right)
\end{split}
}{gcc}
and for the $d-$electron Green's function 
$G_{dd}$ we obtain
\eq{
\begin{split}
 G_{dd}(i\omega_n) = & \frac{1}{i\omega_n +\mu -\epsilon_d    -\Sigma(i\omega_n)} \\
 & +\left[\frac{t_{pd}}{i\omega_n +\mu -\epsilon_d -\Sigma(i\omega_n)}\right]^2 \, G_{pp}(i\omega_n).
\end{split}
}{gdd}
This expression has a transparent physical interpretation: 
there are two processes that a $d-$electron can undergo: 
either remain fluctuating at the local site (first term), 
or fluctuate for some time, 
then jump to the $p$ site and propagate, 
and then return and fluctuate some more time (second term).

\subsection{Numerical solution}\label{subsec:numerical_solution}

For the solution of the associated impurity 
many-body problem (\ref{Seff}), 
one may use a variety of techniques \cite{rmp}. 
Here we shall employ two numerical methods 
which are both a priori exact: 
Hirsch-Fye quantum Monte Carlo (QMC) \cite{hf} and
exact diagonalization (ED). 
The interest of using different techniques is that 
they have complementary range of applicability 
and that they allow for a crosscheck 
of the numerical results. 
The first method, QMC, is a finite temperature calculation 
and is exact in the statistical sense. 
The other method is formulated at $T=0$ 
and rely in a finite size representation 
of the local site environment 
(i.e., the cavity Green's function) 
by a bath of non-interacting atomic sites 
connected to the local impurity. 
In the limit of large number of atomic sites in the bath, 
this approach also becomes a priori exact.
Both methods have already been well documented 
in the literature \cite{rmp}, 
so here we shall limit ourselves 
to briefly provide the relevant technical details.

To implement the QMC,
it is useful to first perform the integration 
on the non-interacting local $p-$site in the action (\ref{self}), 
so that in the many-body problem the interacting $d-$orbital
is the only explicit degree of freedom. 
We then solve the impurity problem 
using the standard Hirsch-Fye QMC algorithm \cite{hf}, 
where the imaginary time interval $[0,\beta]$ is discretized 
in $L$ time-slices of width $\Delta\tau=\beta/L$ 
(where $\beta$ is the inverse temperature).
We set $U\Delta\tau < 1$ to limit the systematic errors 
introduced by the Trotter decomposition. 
The precision of the calculations then basically depends 
on two remaining factors, 
the statistical error and the criterion for the 
convergence of the solution of the DMFT equations.
For the former, we typically perform $10^5$ sweeps. 
When required, we may do up to $10^6$ sweeps, 
such as near the Mott transition, 
or to compute the analytic continuation 
of data to the real axis 
using the maximum entropy method \cite{mem}. 
The quality of the convergence is controlled 
by monitoring the behavior with iteration number of $G_{dd}(i\omega_1)$, 
the  imaginary part of the $d-$electron Green's function 
at the first Matsubara frequency, which shows the largest
variations. 
We stop the DMFT iterations when 
the fluctuations of this quantity become of the order 
of the QMC statistical error 
and remains stable for a few more iterations. 
In generic regions of the parameter space we have studied, 
the solution converge in less than 20-30 iterations, 
but hundreds may be necessary close to a phase boundary.

The ED algorithm is based on 
the representation of the cavity Green's function 
by finite number auxiliary atomic sites \cite{rmp}. 
They conform the ``bath'' or environment of the local impurity.
In general, an infinite number of sites may be required 
to faithfully represent the dynamic environment, 
however, this is not possible to do in practice. 
Therefore, one has to adopt a strategy 
to best represent the environment 
with a finite number of auxiliary sites. 
One may use two different ``geometries'' to represent the bath:
either the ``chain'' geometry, as described 
in Ref.~\onlinecite{edchain} 
or the ``star'' geometry,  as described in 
Refs.~\onlinecite{edstar,goetz}.
In both cases, the effective impurity problem 
consists of a central impurity site,
composed of an explicit $d-$orbital and a $p-$orbital, 
where, by virtue of (\ref{cavity}) only the latter 
is connected to the bath.
In practice, less than 10 sites can be dealt with 
in this method and the ED is performed at $T=0$ 
using the Lanczos technique, 
which is convenient to obtain the Green functions.
In the case of the chain, 
parameters of the auxiliary atomic sites 
can be obtained in terms of a continued fraction expansion 
of the computed Green's functions \cite{edchain,rmp}. 
On the other hand, in the case of the star geometry, 
the parameters are obtained by a $\chi^2$ minimization 
of the difference between the computed local Green function 
and a finite size parametrization of the cavity \cite{edstar,rmp}.

The ED method can be substantially improved 
by supplementing it with 
the Density Matrix Renormalization group 
(DMRG) technique \cite{white_dmrg,dmrgRMP,dmrg_book}.
Several ED-DMRG procedures 
for the solution of the DMFT equations
have been proposed recently \cite{dmrg_book,dmrgRMP,daniel1,gebhard}.
The ED-DMRG method that we use here 
is in essence identical to the ED with the linear chain \cite{daniel1,daniel2}. 
Since the linear geometry is perfectly adapted for the DMRG
procedure, we can ``grow'' the bath 
to contain a higher number of auxiliary sites
with respect to the standard ED \cite{dmrgRMP}. 
In practice, we use up to 40 sites. Nevertheless, we point out that
despite the large number of sites the spectral functions on the real
frequency axis
do not become continuous but still show a discrete multipole structure.
We also observe that the number of poles is roughly similar to the number of sites, but
the spectral weight remains rather concentrated in a relatively
small number of poles, emphasizing the discrete nature of the Green's functions.
However, this does not prevent that the agreement on the Mastubara imaginary
frequency axis is usually excellent (e.g. see Fig.\ref{fig3}).

\section{RESULTS}\label{sec:results}

\subsection{The Mott insulating state}
\label{subsec:mott_insulator}

In this section we shall present our results 
for the Mott-Hubbard regime.
In the Hubbard model, 
the system undergoes a Mott metal-insulator transition 
when the density is $n=1$ 
and the interaction strength $U$ becomes 
of the order of the bandwidth \cite{rmp}.
Here, a similar phenomenon is expected 
as $U$ is increased to a value 
of the order of the effective bandwidth at the Fermi energy 
$\sim t_{pd}^2/\Delta$ and keeping $n_d$ at about $1$. 
As described before, 
the Mott state is realized 
by setting a value of the interaction larger than the effective bandwidth $U>t_{pd}^2/\Delta$ 
and keeping the total occupation $n_{tot}=n_p+n_d=3$.
Similarly to the Hubbard model case, 
in the PAM  there is an on-site Coulomb interaction 
acting on $d-$orbitals.
This interaction punishes the double occupation of $d$ sites, 
and consequently favors the tendency 
to localization and to magnetic moment formation 
of the $d$ electrons. 

Notice that we have chosen in the definition 
of our Hamiltonian (\ref{PAMHam}) and (\ref{Seff}) 
the so called non-magnetic form for the interaction. 
This is motivated by the fact that 
we are interested in the Mott physics 
of a paramagnetic correlated state at $n_d \approx 1$, 
i.e. $n_{d\uparrow}=n_{d\downarrow}\sim 1/2$. 
Therefore the interaction term 
$U\left(n_{d\uparrow}-\tfrac{1}{2}\right)\left(n_{d\downarrow}-\tfrac{1}{2}\right)$ 
approximately cancels at the Hartree-Fock level. 
This allows to obtain some immediate physical insight. 
The cancellation implies that for low values of $U$, 
where first order perturbation holds, 
the interacting density of states of the model 
remains essentially identical as in the non-interacting case 
(see Fig.~\ref{fig2}, upper panels). 
Therefore, the position of the correlated narrow band 
remains approximately fixed at the Fermi energy. 
Since the position remains unrenormalized, 
at higher values of $U$ one would expect that 
the narrow band splits, forming a lower Hubbard band 
and an upper Hubbard band, 
below and above the Fermi energy respectively 
and both carrying half of the spectral 
intensity of the narrow band. 
At the Fermi energy a large charge gap would then open 
and the system becomes a Mott insulator.
\begin{figure}%[!ht]
\centering
\includegraphics[width=8cm,clip]{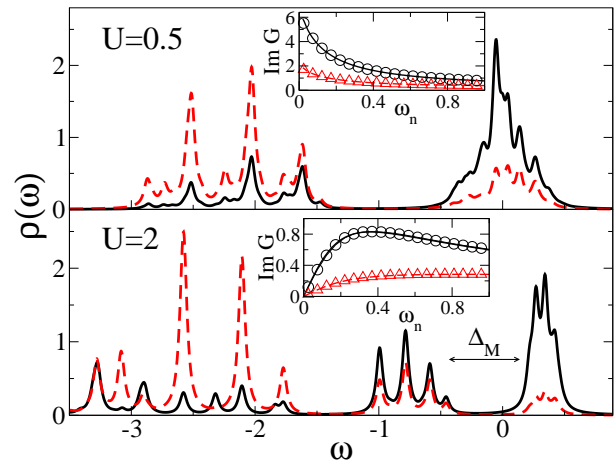}
\caption{
Density of states for the $p-$ and $d-$electrons 
(dashed and solid line) as obtained from 
ED-DMRG calculations (with a 40 sites chain) 
at $\Delta_0=1$, $t_{pd}=0.9$, and $n_{tot}\approx 3$. 
Top panel has $U=0.5$ and $\mu=0.612$. 
Bottom panel has $U=2$ and $\mu=1.029$. 
The arrow indicates the width of the Mott gap $\Delta_{\rm M}$.
Insets: imaginary part of the $p-$ and $d-$electron Green's functions. 
Data are from QMC at $T=1/128$  
(open symbols) and ED-DMRG (lines). 
The finite (zero) value at $\omega_n\rightarrow 0$ shows 
the metallic (insulating) 
character of the solution at $U=0.5$ ($U=2$).
}
\label{fig3}
\end{figure}

\subsubsection{Opening of the Mott gap}\label{subsubsec:opening_mott_gap}

This scenario is in fact borne out 
in the actual model solution that is shown in Fig.~\ref{fig3}. 
The data corresponds to a $T=0$ calculation 
using the ED-DMRG method with 40 sites in the bath.  
The values of the interaction are $U=0.5$ (upper panel) 
and $U=2$ (lower panel). 
This latter value of $U$ 
is sufficient to drive the system 
to the Mott insulating state. In fact one may estimate that the critical $U$ should be about twice the effective bandwidth $t_{pd}^2/\Delta$ (we shall present later on the full phase diagram). 
The insets of the figure contain a comparison 
of the results for the Green's functions in Matsubara frequency 
from ED-DMRG at $T=0$ and QMC at the low temperature $T=1/128$. 
The agreement is very satisfactory. 
The ED-DMRG method also provides 
the propagators in real frequency, 
we thus plot the more intuitive DOS 
in the main panels of the figure. 
In the weakly correlated case (upper panel), 
the DOS resembles the non interacting one: 
at lower energies $-3\lesssim \omega\lesssim-1.5$ 
there is a band with dominant $p-$character, 
while at the Fermi energy 
there is a narrower band with mainly $d-$character.  
In the Mott insulating state shown in the lower panel, 
the DOS consists of three features: 
similarly as before, there is a $p-$like band 
at high (negative) energies. 
However, the main qualitative difference 
is that now the narrow band at $E_F$ is splitted 
in a lower Hubbard band 
and an upper Hubbard band 
respectively below and above the Fermi energy. 
The Mott-Hubbard character of the transition 
in this parameter regime is seen from the fact 
that both lower and upper Hubbard band 
have dominant $d$ character. 
Moreover, one also observes that the $p$ component 
is not negligible, especially in the lower Hubbard band. 

We should also mention that 
the apparent multiple peak structure 
of the main three features appearing in the DOS 
are merely due 
to the discreteness of the finite 
number of sites used to describe the environment 
in the ED technique. 
Though we are using as many as 40 auxiliary sites 
in the environment, 
the discretization effect 
still remains rather noticeable. 
Nevertheless, the splitting of the narrow band 
at the Fermi energy 
with the consequent opening 
of a large Mott gap $\Delta_{\rm M}$ 
that signals the Mott insulator state 
is clearly observed.

\begin{figure}%[!ht]
\centering
\includegraphics[width=8cm,clip]{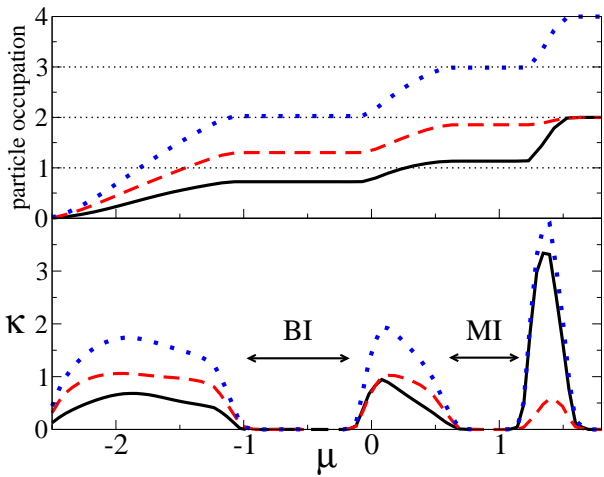}
\caption{Top panel: 
$n_d$ (solid line), $n_p$ (dashed line) and $n_{tot}$ (dotted line), 
as a function of the chemical potential $\mu$, for $U=2$. 
Data are from QMC calculations 
at $\Delta_0=1$, $t_{pd}=0.9$ and $T=1/64$. 
A large Mott gap opens at $U=2$ around $0.7\lesssim \mu\lesssim 1.2$.
Bottom panel: 
charge compressibility, $\kappa=\partial n /\partial \mu$ versus $\mu$ 
for the same model parameter as in the top panel.
}
\label{fig4}
\end{figure}
The transition from metallic to insulating state 
with the consequent opening of the Mott gap, 
can be also observed from the behavior 
of the partial $p$, $d$ and total particle occupation 
$n_{p}$, $n_{d}$ and $n_{tot}$ respectively, 
as a function of the chemical potential.
In Fig.~\ref{fig4} we show these quantities 
and their first derivative $\kappa=\partial n/\partial \mu$ 
proportional to the charge compressibility, for $U=2$.
The plateaux observed in the occupations, 
with the respective vanishing of the compressibility 
for $n_{tot}=0$ and $n_{tot}=4$, 
correspond to the completely empty 
or completely full band insulators. 
The case $n_{tot}=2$ corresponds to 
the hybridization band insulator, 
similar to the one already discussed 
in the non-interacting case. 
However, in contrast to the results 
for the non-interacting case, 
the strong value of the interaction 
creates additional plateaux in the $n(\mu)$ curves.
The new plateaux occur when 
the total number of particles is exactly $n_{tot}=3$, 
which is odd and signals the Mott state. 

An important aspect that we should mention is that 
the Mott insulator state occurs where the correlated $d$ site 
has an occupation close to one, but not exactly unity. 
This can be seen is Fig.~\ref{fig4} (top panel), 
where the Mott plateau occurs at $n_d=1+\nu$ (and $n_p=2-\nu$), 
with $\nu\approx0.13$.  
The specific value of $\nu$ depends on the hybridization 
and therefore one can consider it 
as a measure of the mix-valence character 
of the Mott insulating state. 
It is the {\sl total} number of particles exactly 
equal to three (or one hole) what is required 
for the onset of the Mott insulator state. 
This implies that in this model 
where the mixed $p-d$ valence is explicitly included 
through the hybridization, the Mott localization occurs 
for a ``composite'' object 
which has a mixed $p$ and $d$ character.

\subsubsection{Size of the Mott gap}
\label{subsubsec:size_mott_gap}

\begin{figure}%[!ht]
\centering
\includegraphics[width=8cm,clip]{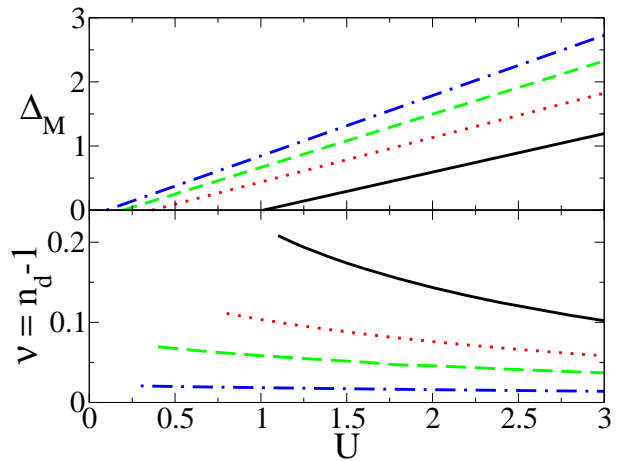}
\caption{Top panel: 
size of the Mott gap $\Delta_{M}$ as a function of $U$, 
for different position of the $p$ electron band 
$\epsilon_p=-6,-3,-2,-1$ (top to bottom). 
Lower panel: 
mix-valence character $\nu=n_d-1$ as a function of $U$, 
for the same position of the $p$ band 
as in the upper panel (bottom to top). 
The results are obtained from ED.
}
\label{fig5}
\end{figure}
The size of the Mott gap is naively expected 
of be of order $U$, 
since it should mostly reflect the energetic cost 
to doubly occupy the $d$ orbitals.
However our results show 
that the size of the Mott gap $\Delta_{\rm M}$ 
may be substantially smaller than the bare value $U$.
In Fig.~\ref{fig5} we plot $\Delta_{\rm M}$ 
as a function of $U$, 
for several values of the bare position 
of the $p-$band $\epsilon_p$, 
which amount to increase the charge transfer energy $\Delta_0$. 
As the energy of the $p-$orbitals is shifted down 
to larger (negative) energies, 
the effective bandwidth of the narrow band 
at the Fermi energy decreases. 
In addition, 
the $p-$electron band becomes essentially full 
with $n_p \to 2$ as $\epsilon_p \to -\infty$ 
(and keeping $t_{pd}$ fixed). 
This implies a decrease in 
the mix-valence character of the electrons 
at the Fermi energy $\nu$,  
as shown in the lower panel of Fig.~\ref{fig5}.
In this limit 
the size of the Mott gap approaches 
the ``bare'' value $\Delta_{\rm M} \approx U$. 
However, it is interesting to observe that 
the smaller values of $\epsilon_p$ lead to a substantial 
renormalization of the size of the expected Mott gap.
This effect can be thought as due to 
an effective screening that the $p-$electrons provide, or, 
in more naive terms, because the electrons 
only ``feel'' the repulsive term $U$ during the time 
they spend on the $d-$orbital, 
but not when they visit the $p$ site. 
So as the mixed $p-d$ character is increased, 
the effect of the $U$ is renormalized downwards.

\begin{figure}%[!ht]
\centering
\includegraphics[width=8cm,clip]{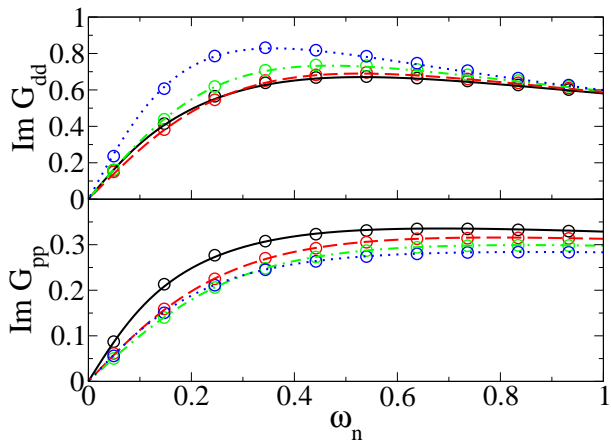}
\caption{Top panel: 
imaginary part of the $d-$electron Green's functions 
in the Mott state, 
from ED-DMRG (for a large finite clusters of $30$ sites) 
at $T=0$ and $\mu=0.729, 0.829, 0.929, 1.029$ 
(solid, dashed, dotted-dashed, dotted line respectively). 
Circles are the same quantity from QMC at $T=1/64$.
Lower panel: 
imaginary part of the $p-$electron Green's functions; 
all model parameters are the same as in the top panel.
}
\label{fig6}
\end{figure}

For completeness we show the behavior 
of the imaginary part of the Green's functions 
at low Matsubara frequency. 
In Fig.~\ref{fig6} we present numerical results 
for several insulating states 
obtained varying the chemical potential within the Mott plateau. 
The data were obtained with both QMC and ED-DMRG, 
so also serve to illustrate the good agreement 
between the two methods.
Note that the imaginary part 
of both the $p$ and the $d$ electron Green's functions 
goes to zero at $\omega_n \rightarrow 0$. 
By analytic continuation this implies that 
the $p$ and $d$ electron DOS vanish at the Fermi level, 
consistent with the insulating character of the solutions.
Accordingly, when the chemical potential $\mu$ 
is varied within the Mott plateau, 
the $p-$ and $d-$electron DOS simply experience 
a rigid shift in energy. 
However, as we shall see later, 
there are dramatic changes in the DOS lineshapes 
as the system turns metallic upon doping.

\subsection{Doped Mott insulator}\label{subsec:doped_ins}

So far we have shown that 
the model does have a Mott insulator state 
which similarly as in the Hubbard model case 
develop two incoherent Hubbard bands 
above and below the Fermi energy. 
However, unlike in the Hubbard model case, the size of the gap, 
i.e. the separation between the Hubbard bands, 
may be substantially smaller than $U$ 
if the hybridization is relatively high.
In the following section we shall proceed to dope 
this Mott insulator with $\delta$ carriers, 
with $\delta=n_{tot}-3$. 
As was already reported 
in Ref.~\onlinecite{sar}, 
we will observe that the insulator to metal transitions 
that can be obtained by either particle, $\delta > 0$, 
or hole doping, $\delta < 0$, 
are qualitatively different.
The former will essentially reproduce the known scenario 
for the Mott MIT that is realized in the DMFT solution 
of one band Hubbard model \cite{rmp,sahana,daniel2,werner}.  
This was to be expected 
since we have tuned the parameters of the model to the regime 
where the identification of the low energy physics of the PAM 
and the one band Hubbard model, was expected to hold 
\cite{pruschke1}.
However we shall see that, rather surprisingly, 
the hole doping insulator to metal transition 
bears out a qualitatively different scenario.

\subsubsection{particle doping ($\delta > 0$)}
\label{subsubsec:particle_doping}

In this section we shall first describe the MIT 
driven by particle doping and demonstrate that 
it realizes the same first order transition scenario 
as the one in the single band Hubbard model.  
The metallization of the Mott insulator 
is most directly seen from the changes that take place 
in the density of states.
\begin{figure*}[!ht]
\centering
\includegraphics[clip=true,width=0.95\textwidth]{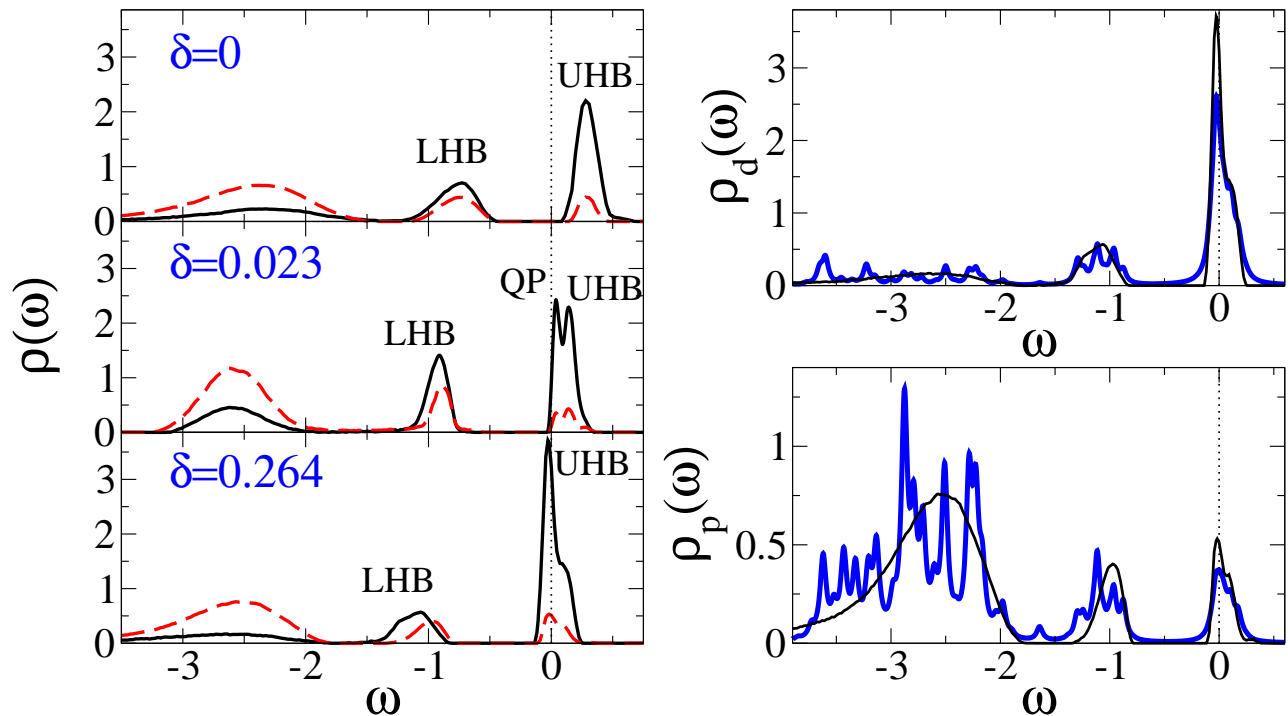}
\caption{
Density of states for the $p$ and $d$ electrons 
(dashed and solid line) 
at $U=2$, $\Delta_0=1$, $t_{pd}=0.9$. 
Left: DOS from analytically continued QMC data at $T=1/64$.
Top panel has $\mu=1.079$, corresponding to a Mott insulating state ($\delta=0$).
Middle panel has $\mu=1.244$, corresponding to small doping $\delta=0.023$.
Bottom panel has $\mu=1.317$, corresponding to high doping $\delta\approx0.26$.
Right panels: comparison of DOS from ED-DMRG and QMC data (thick and thin line 
respectively) at $\delta=0.264$.
}
\label{fig7}
\end{figure*}
In Fig.~\ref{fig7} we show the evolution of that quantity 
for the $p-$ and $d-$electron components 
as a function of doping. 
The data are obtained from analytical continuation of 
QMC results at $U=2$ and low temperature. 
To compute the analytical continuation using the MEM, 
we used over $10^6$ sweeps, so to minimize the uncertainty due to statistical errors.
In the left top panel we see the DOS for the case where 
the chemical potential $\mu$ is close to the upper energy edge 
of the Mott gap, therefore still in the insulator state with $\delta=0$. 
The left bottom panel shows the metallic state that is obtained 
when the chemical potential enters the upper Hubbard band, 
doping the parent Mott insulator with particles. 
We observe a broad peak at the Fermi energy 
and a strong transfer of spectral weight 
from the lower to the upper Hubbard band. 
The $d$ orbital character remains dominant in the DOS upon doping.
In the left middle panel we show 
the DOS in the region of small doping, very near the transition. 
The data reveal that both the $p$ and $d$ electron components 
of the DOS show a narrow quasiparticle peak at the Fermi energy, 
flanked by the incoherent upper Hubbard band at higher energies. 
This coherent peak 
carries a small fraction of the spectral intensity 
which is of order $\delta$. 
In addition, from the enhancement of the slope of the self-energy, 
one observes that the quasiparticles acquire a heavy mass. 
All these features are consistent with the MIT scenario 
found in the one band Hubbard model.

The right panels of Fig.~\ref{fig7} show a comparison of QMC and ED-DMRG results 
for the DOS at high doping. 
The discrete peaks in the ED-DMRG data 
are due to the finite number of sites in the effective bath. 
Nevertheless, the agreement in the distribution of the spectral weight of the two methods 
is very satisfactory. 
The comparison of the data also allows for a non-trivial benchmark of the numerical results and is a useful illustration of the advantages and disadvantages of the different numerical techniques. ED-DMRG is exact but has
the drawback of discrete poles structures that persist even for many effective
atoms in the bath. QMC, in contrast, produce smooth spectra but their numerical precision cannot be guaranteed due to the uncertainties in the process of analytic continuation of data to the real frequency axis. 

\begin{figure}%[!ht]
\centering
\includegraphics[width=8cm,clip]{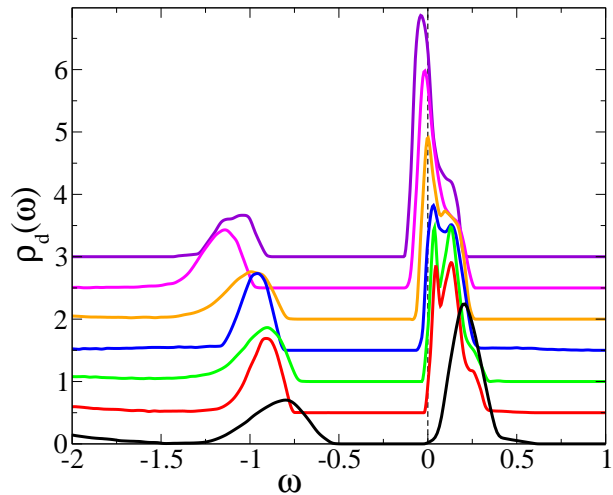}
\caption{Density of states for the $d$ electrons from QMC data at 
$U=2$, $\Delta_0=1$, $t_{pd}=0.9$, $T=1/64$. 
The particle doping is $\delta=0$, $0.01$, $0.02$, $0.04$, $0.11$, $0.21$, $0.32$ 
(bottom to top).
}
\label{fig7b}
\end{figure}
In Fig.~\ref{fig7b} we present the $d-$electron density of states $\rho_d(\omega)$ for several values of particle doping. The data show
the presence of the quasiparticle peak evolving in between the two Hubbard
bands. One may see that though the data show a reasonable systematic evolution some details are beyond the precision of the technique, such as the precise
size of the gap and the form of the bandedges.

\begin{figure}%[!ht]
\centering
\includegraphics[width=8cm,clip]{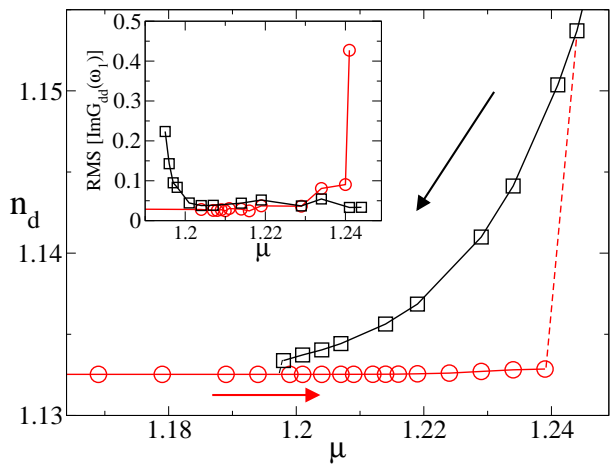}
\caption{
Hysteresis loop 
for the $n_d(\mu)$ curve at the upper energy edge 
of the Mott insulating state. 
Data are obtained from QMC 
for $T=1/64$ at $U=2$, $\Delta_0=1$, $t_{pd}=0.9$. 
The arrows indicate the solution 
obtained by following the insulating (circles) 
and the metallic solution (squares). 
Inset: 
root mean square (RMS) of the imaginary part 
of the $d-$electron Green's function value 
at the first Matsubara frequency 
as a function of the chemical potential. 
It shows the enhancement of the fluctuation 
near the two critical values of chemical potential, 
where the continued solution ceases to exist.
}
\label{fig8}
\end{figure}

The most substantial confirmation 
that the particle doping driven MIT scenario
in the PAM and in the HM are in fact qualitatively analogous 
comes, however, from the observation of the hysteresis effect 
in the particle number $n(\mu)$ curve.
The hysteresis is a hallmark of the first order nature 
of this doping driven transition
and it was observed and studied in detail 
in the Hubbard model \cite{rkz,gkrauth2,oudovenko,sahana,werner}.
We also find it here in the PAM, 
and it is most clearly appreciated in the behavior 
of $n_d$ versus $\mu$. 
There is a strong dependence 
of the $n_d(\mu)$ curves as the temperature is lowered, 
signaling strong correlation effects been active with very low
energies. 
In Fig.~\ref{fig8} we show the occupation of the $d$ electrons 
as a function of the chemical potential $\mu$, 
obtained from QMC.
The main panel shows a detail of the QMC data 
at the low temperature $T=1/64$, 
where the hysteresis loop can be clearly seen. 
These results were verified using the $T=0$ ED technique. 

In order to observe the hysteresis cycle, 
we use as a seed for the iterative procedure 
the converged solution from the previous set of parameters 
\cite{rkz,sahana}.
Thus, the solutions can be continuously ``followed'' 
in parameter space, 
until it shows a sudden jump. 
The discontinuous jump occur at the approximate
position of the spinodal lines
\cite{chaikin}.
Similarly as in previous studies on the Hubbard model \cite{rkz}, 
the hysteresis defines a region of parameters 
where two solutions of the DMFT equations 
can be stabilized. 
The true physical transition should occur 
where the free energy of the solutions cross. 
The precise determination of that line 
is beyond the scope of our present study 
and probably would require further refinement
of the numerical techniques 
due to the very low energy scales involved.

One key point that provides further support to our results 
is the critical slowing down phenomenon 
observed in the QMC calculation 
at the phase boundaries of the coexistence region 
\cite{oudovenko,sahana}. 
This phenomenon is characterized by 
an enhancement of the number of iterations
required to achieve self-consistency 
and also by an enhancement of the statistical Monte Carlo fluctuations 
that reveal the shallowness of the energy landscape 
when two solutions merge. 
The root mean square deviation of the lowest frequency component
of the $d-$electron Green's function 
is plotted in the inset of Fig.~\ref{fig8} 
as a function of $\mu$ 
for the low temperature $T=1/64$. 
As we approach the phase boundaries of the coexistence region, 
we can see that the root mean square deviation increases. 
Starting with an insulating solution, 
if we increase the chemical potential, the RMS increases 
until we reach a critical value of $\mu$, 
where the insulating solution disappears (open circles).  
Similarly, starting form a metallic solution at low doping, 
if we reduce the chemical potential, the RMS grows 
until a critical value of $\mu$, 
where the metallic solution disappears (open squares). 

\begin{figure}%[!ht]
\centering
\includegraphics[width=8cm,clip]{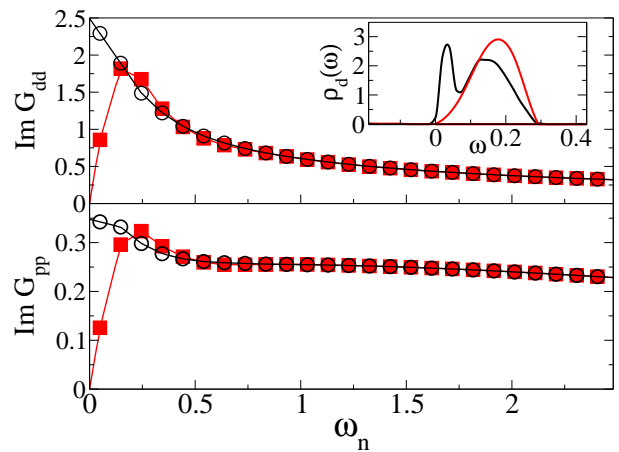}
\caption{Main panel: 
imaginary part of the $d-$electrons Green's function 
as a function of the Matsubara frequency 
for $\Delta_0=1$, $t_{pd}=0.9$, $U=2$, $T=1/64$ 
and $\mu=1.234$, corresponding to tiny particle doping 
$\delta=n_{tot}-3\approx0.01$. 
Open symbols correspond to the metallic solution, 
while full symbols to the insulating one.
Inset: 
the correspondant $d-$electron density of states 
$\rho_d(\omega)$. 
Thick line is the metallic solution, 
thin line is the insulating one.
}
\label{fig9}
\end{figure}

Evidently, the coexistence of solutions can also be observed 
from the behavior of other quantities, 
such as the double occupancy, 
or the low frequency part of the Green's functions. 
The latter is shown in Fig.~\ref{fig9} 
for both $p-$ and $d-$electron components.

These results support the claim that
the MIT scenario for $\delta>0$ in the PAM  
is completely analogous to the one found 
in the Hubbard model investigations 
\cite{sahana,daniel2,werner}.

\subsubsection{hole doping ($\delta <0$)}
\label{subsubsec:hole_doping}

\begin{figure*}[!ht]
\centering
\includegraphics[clip=true,width=0.95\textwidth]{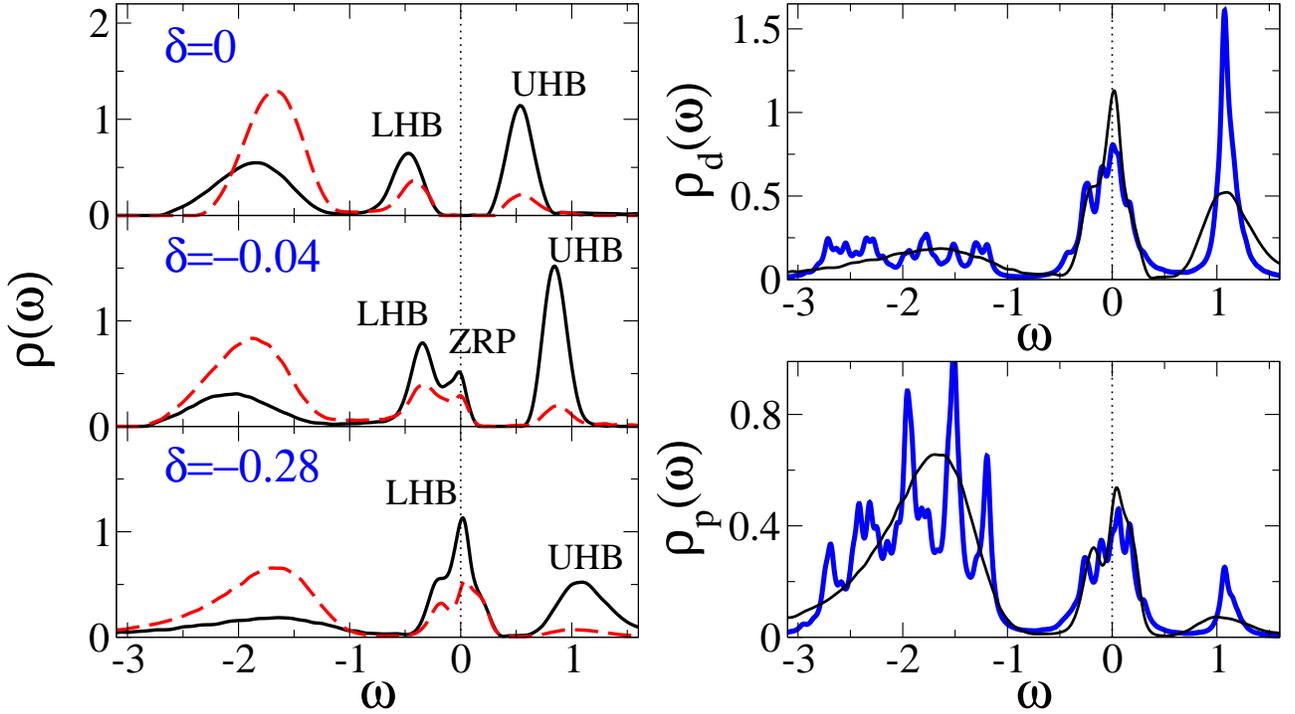}
\caption{
Density of states for the $p-$electrons (dashed line) 
and $d-$electrons (solid line) 
for $U=2$, $\Delta_0=1$, $t_{pd}=0.9$.
Left: DOS from analytically continued QMC data at $T=1/64$.
Top panel has $\mu=0.829$, corresponding to a Mott insulating state ($\delta=0$). 
Middle panel has $\mu=0.529$, corresponding to small doping 
$\delta=n_{tot}-3=-0.04$.
Bottom panel has $\mu=0.329$, corresponding to high doping $\delta=-0.28$.
The QMC data show the appearance 
of a broad quasi-particle peak (ZRP) at the Fermi level, 
in addition to the lower and upper Hubbard bands (LHB, UHB) at high energies.
Right panels: comparison of DOS from ED-DMRG and QMC data (thick and thin line 
respectively) at the heavy doping $\delta=-0.28$.
The ED-DMRG data are obtained using an environment of 30 sites.
}
\label{fig10}
\end{figure*}
In the previous section we showed that 
upon particle doping, $\delta > 0$, 
the Mott MIT in the PAM realizes the same physical scenario 
as the one observed in the single band Hubbard model. 
While in the latter the particle-hole symmetry 
implies an identical transition for $\delta<0$,
we shall see that this is not the case in the PAM.
A key point to appreciate is that 
we shall keep all model parameters fixed, 
with the obvious exception of the chemical potential, 
which controls the occupation. 
Therefore, if the chosen model parameters led, for $\delta>0$, 
to the identification of the PAM with the HM physics, 
then one may also expect that 
this would be the case for $\delta < 0$ as well. 
Rather surprisingly this turns not to be the case \cite{sar}. 
In this section we shall describe the main physical behavior
of the model for the hole doping driven MIT, 
and in the next we shall argue 
about the origin of this unexpected result. 

We begin by showing 
the effect of hole doping in the DOS. 
In Fig.\ref{fig10} we plot the change 
in the $p$ and $d$ components of the DOS 
as the system evolves from Mott insulator 
to a hole-doped metallic state (left panels, top to bottom). 
In the insulator state we observe that
the chemical potential is located within the correlation gap, 
and the lower and upper Hubbard bands 
can be well appreciated. 
As we already discussed before, in this Mott insulator state 
the DOS at low frequencies has mostly $d$ electron character, 
since the $d$ orbital was initially located at the Fermi energy. 
Upon hole doping, 
the chemical potential moves within the lower Hubbard band. 
The metallization produces a wide and strong quasiparticle peak 
at the Fermi energy. 
At small doping, 
the low frequency part of the spectrum has 
a characteristic three peaks structure: 
the lower Hubbard band around $\omega\sim-0.4$, 
the quasiparticle peak that crosses the Fermi level, 
and the upper Hubbard band at $\omega\sim0.8$  
An interesting aspect to appreciate is that 
the quasiparticle peak has, as before, a  
larger $d$-character, but, in addition, now it 
also has a substantial $p$-electron component. 
We shall see that this aspect will be consistent with the interpretation
of the quasiparticle peak now emerging not from mere delocalization of
$d$-electrons as before, but from the delocalization of a composite object
that involves a $p$-hole and a $d$-electron spin. 

We can also observe a transfer of spectral weight, 
with an increase of the 
relative intensity of the lower Hubbard band plus quasiparticle, 
at the expense of a decrease of the upper Hubbard band.
The structure that is seen below $\omega \approx -1.5$ 
corresponds to the fully filled band 
which remains with predominantly $p-$electron character. 
The results shown were obtained 
with high quality QMC data (at least $10^6$ sweeps to reduce
the statistical errors) 
and performed the analytic continuation to the real axis 
using the maximum entropy method \cite{mem}.

In the right panels of Fig.~\ref{fig10} we present 
a comparison of QMC and ED-DMRG results for the DOS at heavy doping $\delta=-0.28$. 
Similarly as before (cf. Fig.~\ref{fig7}), 
the apparent multi-peak substructures of the ED-DMRG results are not physical 
and only due to the discrete number of poles that result 
from a finite number of atomic sites in the auxiliary bath. 
Compared to the standard ED, this method produces much smoother spectra
with greater detail due to the dramatic increase in the number of poles.
However, as is also evident from the data, the discreteness due the
finite size representation of the bath remains a shortcoming of the method. 
The overall comparison of the two methods remains, nevertheless, very satisfactory and serves to illustrate their relative advantages and disadvantages.  

\begin{figure}%[!ht]
\centering
\includegraphics[width=8cm,clip]{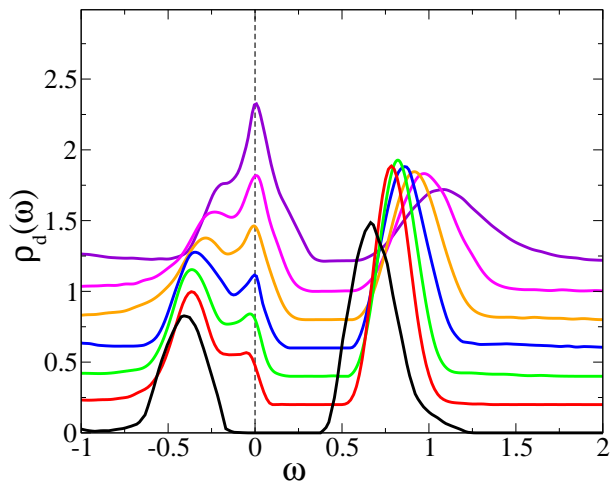}
\caption{Density of states for $d$ electrons for several values of hole doping, 
from analytically continued QMC data 
at $U=2$, $\Delta_0=1$, $t_{pd}=0.9$, $T=1/64$. 
From bottom to top, the hole doping is 
$\delta=0$, $-0.01$, $-0.03$, $-0.04$, $-0.09$, $-0.15$, $-0.28$. 
}
\label{fig10b}
\end{figure}
To complete this study, we show in Fig.~\ref{fig10b} the detailed evolution of the quasi-particle peak at the Fermi level plus the Hubbard bands as a function of doping. The data correspond to the $d-$electron density of states obtained from analytic continuation of QMC data. 

The emergence of a quasiparticle peak 
at the Fermi energy for the hole doped case 
may seem, at first sight, 
similar to the metallic state obtained 
from particle doping.
However, this will not turn to be the case \cite{sar,asr08}
and the physical origin of the two quasiparticle excitations
will be shown to be qualitatively different. 
The reason for this unexpected asymmetry
will be discussed in the next section.

\begin{figure}%[!ht]
\centering
\includegraphics[width=8cm,clip]{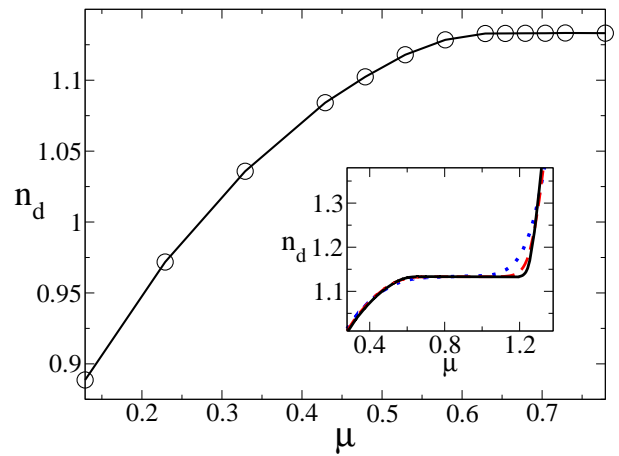}
\caption{
Partial occupation of the $d$ electrons, $n_d$, 
as a function of the chemical potential $\mu$ 
at $U=2$, $\Delta_0=1$, $t_{pd}=0.9$. 
Data are obtained from QMC calculations. 
Main panel has $T=1/128$. 
No trace of coexistent solutions are found. 
Inset has $T=1/16$, $1/32$, $T=1/64$ 
(dotted, dashed and full lines). 
Notice the strong temperature dependence 
on the particle doped side of the transition 
(where only one branch of the hysteresis cycle at $T=1/64$ is shown), 
and the much less temperature dependence 
on the hole doped side of the transition.
}
\label{fig11}
\end{figure}

In order to fully underpin 
the nature of the order of the hole doped transition, 
we now look for hysteresis effects. 
Thus, as we did before, 
we continuously follow the solutions in parameter space. 
First, we start from the insulator 
and lower the chemical potential 
till we obtain a significantly doped metal; 
and then, we start from the metal 
and increase $\mu$ until we reach again the insulator. 
Neither our QMC numerical simulations 
nor the ED-DMRG studies 
showed any indication of hysteresis effects. 
The QMC data down to $T=1/128$ 
are shown in Fig.~\ref{fig11}.
Compared to the results for particle doping (inset), 
the present ones show 
a negligible temperature dependence.
Thus, up to our current numerical capacity 
we have to conclude that 
the metal-insulator transition 
in the hole doped case 
is of {\sl second order}.
As no evidence of coexistence has been found, neither in finite or zero
temperature calculations, we are led to the conclusion that the
hole doping driven transition at $T=0$ is continuous, i.e. has a second order character.
Thus, it is qualitatively different 
from the particle doped case, and, 
consequently, also qualitatively different 
from the Hubbard model scenario.
Of course, we cannot rule out 
the eventual existence of tiny energy scales (which may
modify our proposed scenario) which remain beyond the numerical 
precision of our methodology. 
This issue cannot be resolved 
nor by ED or ED-DMRG data 
due to the finite frequency cut-off 
set by the finite size of the clusters 
to diagonalize. 
Resolving this issue would probably require NRG study. 
However this is not fully clear 
since NRG method requires a good separation of energy scales, 
which is not the case here.

\section{DISCUSSION}\label{sec:discussion}

Our results for the doping driven MIT, that arise 
from either particle or hole doping of the Mott insulator, 
show a qualitative asymmetry, 
thus questioning the expected mapping of the PAM 
onto the Hubbard model. 
In this section we shall address in more detail this issue 
from the perspective 
of the physical nature of the two MIT 
taking place in the PAM.

\subsection{Phase diagram}

\begin{figure}%[!ht]
\centering
\includegraphics[width=8cm,clip]{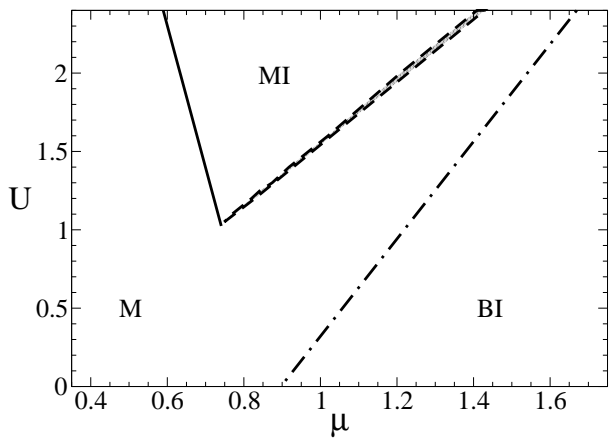}
\caption{
Phase diagram of the PAM in the $U-\mu$ plane 
that maps the Mott insulator (MI) 
and its boundaries with the metallic state (M). 
The boundary between the metal 
and the band insulator (BI) 
at $n_{tot}=4$ is also shown.
The boundary lines are for 
$T=0$ ED star geometry calculations 
at $\Delta_0=1$ and $t_{pd}=0.9$. 
The dashed line denotes the region of the parameters 
where there is coexistence between two solutions, 
one metallic-like and the one insulating-like. 
}
\label{fig12}
\end{figure}
To begin, we map out the phase diagram 
in $U-\mu$ parameter space 
to explore the respective ranges 
of the first and the second order transitions. 
Our results  are summarized in Fig.~\ref{fig12}, 
that was constructed using finite $T$ quantum Monte Carlo 
and $T=0$ ED data. The latter were obtained using the ``star-bath'' geometry, 
which we found better suited \cite{rmp,edstar} 
than the ``chain-bath'' geometry 
to study the possible coexistence 
between the metallic and the insulating regions.

The phase diagram in the $U-\mu$ plane 
shows a central V-shaped Mott insulator region 
with the correlated metallic phases 
for particle and hole doping, 
respectively to the right and to the left. 
A threshold value for the strength of the interaction $U$ 
(tip of the V-shape boundary) 
is required to obtain a Mott insulator state. 
This threshold depends on the value of the
``bandstructure'' parameters $\Delta_0$ and $t_{pd}$.
This feature is analogous  
to the existence of critical value 
of the ratio $U/D$ in the one band Hubbard model \cite{rmp}.  

The central V-shaped Mott insulating region 
shows a remarkable asymmetry  
comparing the hole and the electron sides.
In contrast, in the one band Hubbard model, 
due to the particle-hole symmetry, 
the V-shape onset of the Mott insulator 
is symmetric respect to the tip 
and behaves as $U_c\sim U_{c1} \pm 2\mu$, 
where $U_{c1}$ is the value of the interaction 
at which the insulator disappears \cite{rmp}. 
In the PAM, the transition line 
to the particle doped side behaves as well like 
$U_c\sim {\rm const}+2\mu$. 
However, the transition line 
to the hole doped metal 
is almost vertical.

We also mapped out the order of the transition 
along the boundary lines. 
Upon hole doping, 
both QMC calculations 
down to the low temperature $T=1/128$, 
and ED calculations at $T=0$ 
show no trace of coexistent solutions
along all the transition line, 
indicating a second order transition. 
On the other hand, upon particle doping, 
the dashed line in the phase diagram 
displays the region of the parameter space 
where, for sufficient low temperatures, 
an insulating state 
coexists with a metallic state. 
In fact within ED method at $T=0$ 
we find that the DMFT equations 
have two different solutions 
all along the transition line. 
Within QMC, and down to the low temperature $T=1/128$, 
we found a coexistence region only 
for a sufficiently large strength 
of the repulsive interaction $U$ 
(approximately $U\approx 2$ 
for our specific choice of parameters).
However, for smaller values of $U$, 
the phase boundary 
remains strongly dependent on temperature. 
This suggests that the temperature 
below which there is a coexistence 
between metallic and insulating state  
decreases rapidly approaching the tip of the V-shape. 
This should be expected, 
since the two boundary lines, 
to holes and particle doped metal, 
eventually merge at the tip. 
Similarly, we also expect the $T=0$ width of the
coexistent region to become narrower as one
approaches the tip of the V. However, obtaining detailed
and precise numerical data there turned out to be beyond 
our methods' capabilities.

We note that at the Mott insulator state 
($\delta=0$) the occupation of the conduction $p-$electrons
is almost saturated at $n_p = 2-\nu$ (with $\nu \approx 0.1$). 
On the other hand, the occupation 
of the non dispersive $d-$electrons 
which carry strong magnetic moments 
due to the on-site repulsion 
is close to one, with $n_d = 1+\nu$ (cf Fig.~\ref{fig5}). 
Therefore, there is only a small number $\sim \nu$ of $p-$holes 
available to screen a number of order one of $d$ magnetic moments.
Therefore, a natural issue to consider is 
whether the physics associated to the ``exhaustion problem'' 
of Nozi\`eres \cite{nozieres1,nozieres2,gkepjb} 
(see also related works 
on the Kondo lattice model \cite{burdin,costi} 
and on the PAM \cite{tjf1,protracted,pruschke}) 
may play a role in the different transitions 
at $\delta$ greater or smaller than $0$.

An important point to realize is, however, 
that the exhaustion situation 
is even more extreme {\it on the particle side 
than on the hole-doped side}.
In fact, while the number of $d-$electron remain always 
of order one on both sides, the number of available $p-$holes 
is substantially smaller 
for $\delta>0$ with respect to the $\delta<0$ case. 
Nevertheless, the Hubbard-like first-order transition scenario
takes place only on the particle-doped side. 
In other words, the PAM metal-insulator transition scenario 
is analogous to the one in the Hubbard model 
when the PAM is even deeper in the exhaustion limit 
($\delta>0$, $n_{p}\approx 2$).
This implies that while exhaustion should play a role, 
it is not obviously responsible for the failure of the mapping 
of the PAM onto the HM for $\delta<0$.

\subsection{Nature of doped carriers}
\begin{figure}%[!ht]
\centering
\includegraphics[width=8cm,clip]{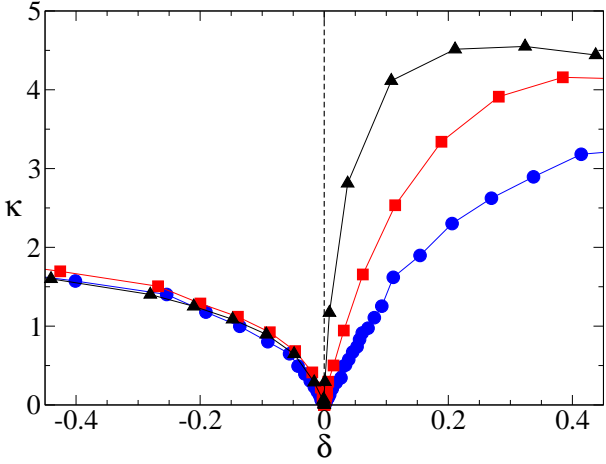}
\caption{
$\kappa=\partial n_{tot}/\partial\mu$ 
as a function of doping $\delta$, 
for several values of the temperatures $T=1/16,1/32,1/64$ 
(circles, squares, and triangles respectively). 
Data are obtained from QMC calculations 
at $\Delta_0=1$, $t_{pd}=0.9$, $U=2$.
On the particle doped side, 
we followed the metallic solution. 
}
\label{fig13}
\end{figure}

We now address the issue of the physical nature
of the metallic states in this system. 
Although the low frequency part of the DOS of both, 
particle and hole doped metals, 
have a similar three peaks structure 
with dominant $d$ character, 
they realize physically different states. 
For instance, their charge compressibility 
has a very contrasting temperature dependence. 
In Fig.~\ref{fig13} we show 
the derivative of the total occupation 
with respect to the chemical potential, 
$\kappa=\partial n_{tot}/\partial\mu$ 
(an observable proportional to the compressibility) 
as a function of doping. 
In the Mott insulator ($\delta=0$) 
$\kappa$ is zero, 
indicating the incompressible Mott state. 
As expected, upon doping, $k$ increases, 
indicating that the system becomes compressible. 
For small particle doping, 
the compressibility rapidly increases with the temperature, similarly
as in the one band Hubbard model case. 
Upon hole doping, $\kappa$ is much less dependent on temperature.

In the PAM the physics of the metallic states 
is usually discussed in terms of the screening between 
the $p$ conduction electrons (or holes) and 
the magnetic moment of the local $d-$electrons. 
In fact, at each lattice site a $p-$ and a $d-$electron 
may form a local singlet, 
which is the underlying idea 
in the argument of Zhang and Rice (ZR) in the context 
of high temperature cuprate superconductors \cite{zr}. 
However, if on the particle side of the Mott MIT 
we have argued that 
there are essentially no available holes, 
then the question is, what is screening the $d-$moments 
so to produce a normal (but heavy) Fermi liquid metal, 
analogous to the one in the doped Mott insulator 
in the Hubbard model? 

\begin{figure}%[!ht]
\centering
 \subfigure[]{\label{fig14a}
\includegraphics[width=7cm]{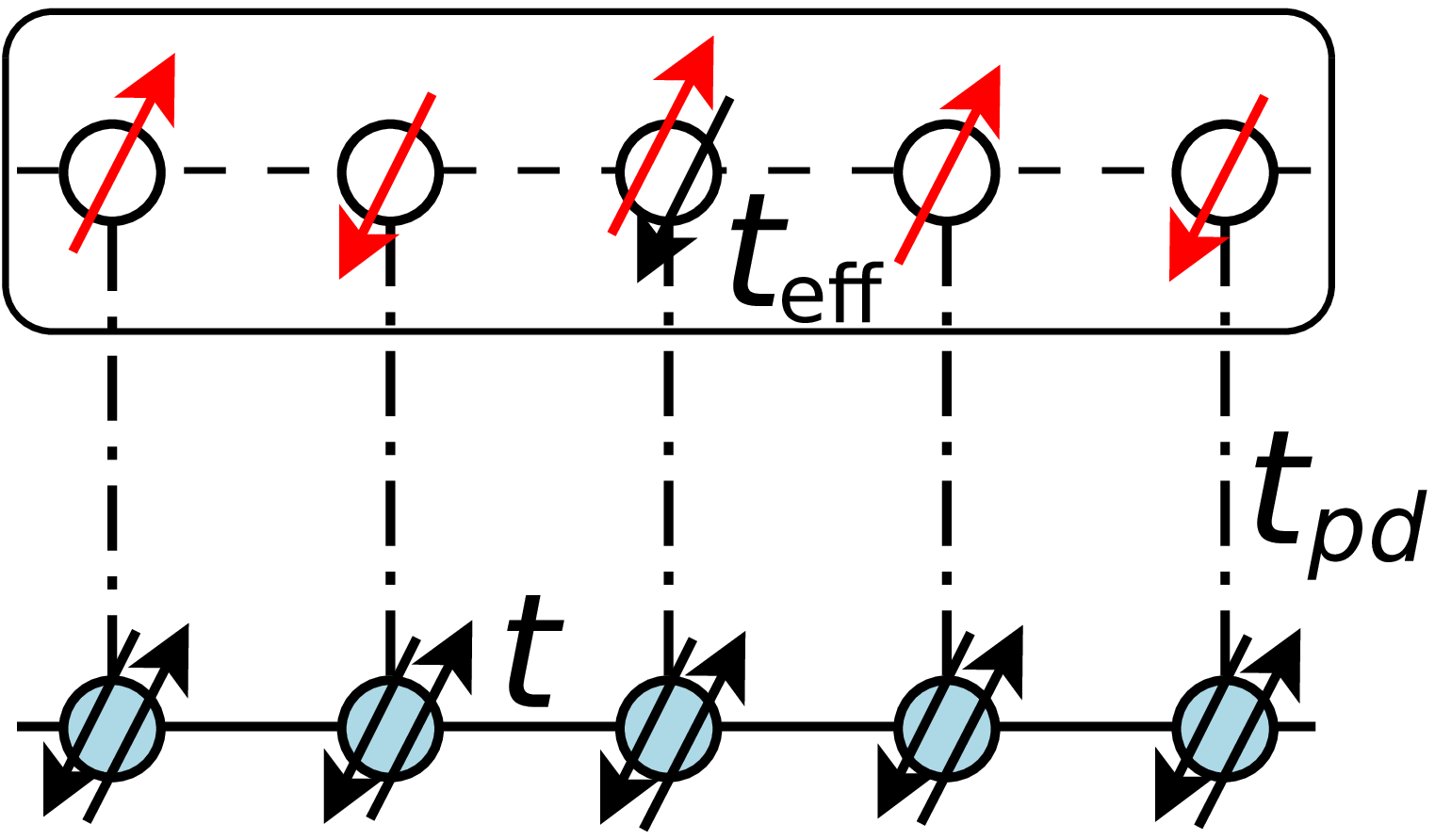}}

 \subfigure[]{\label{fig14b}
\includegraphics[width=7cm]{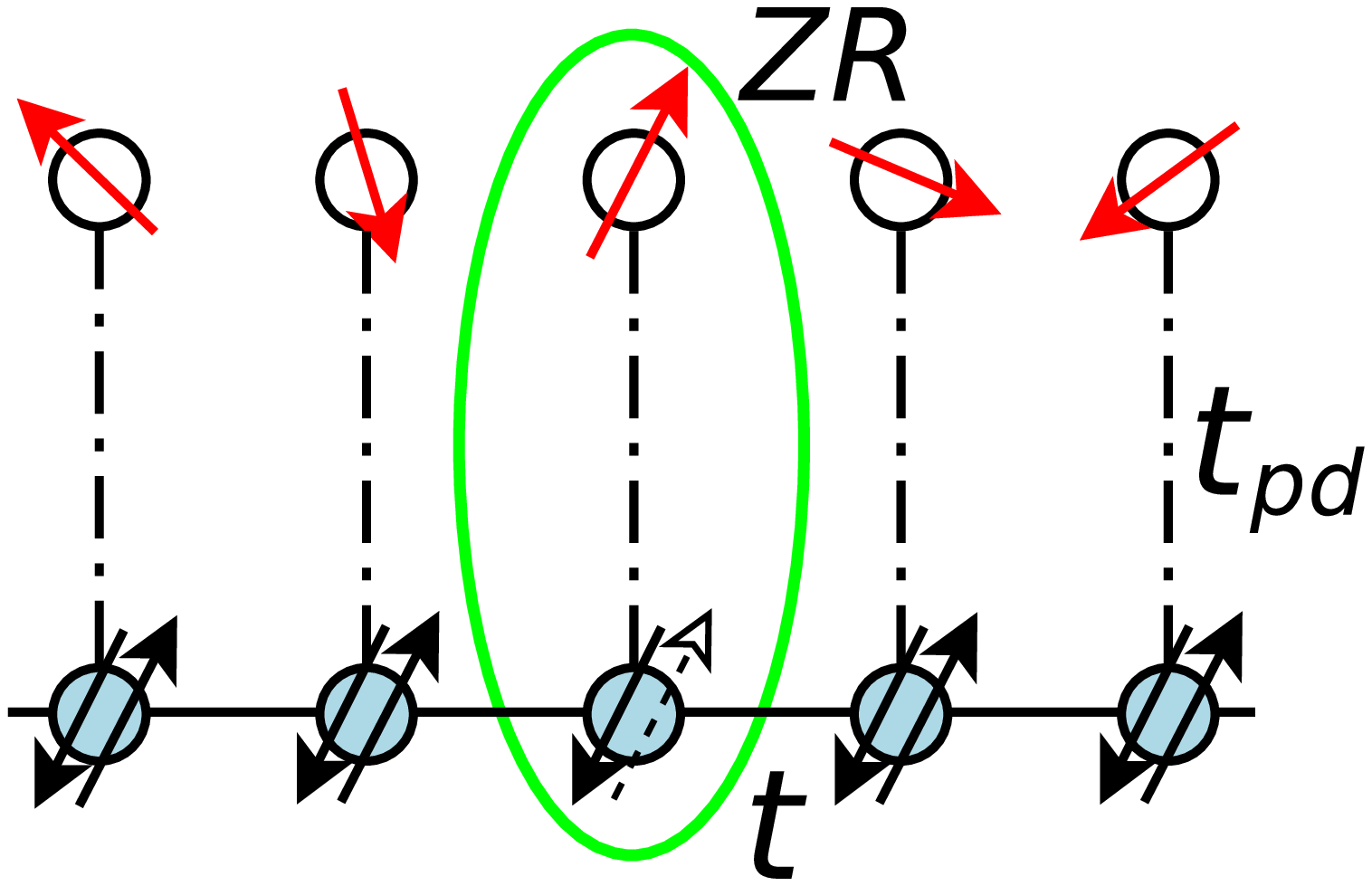}}
\caption{Schematic representation of the different type of carriers that
occur upon either (a) particle ($\delta > 0$), or  (b) hole ($\delta<0$)
doping of the Mott state.}
\label{fig14}
\end{figure}
The answer to this question is that, 
similarly to the one band Hubbard model case, 
the $d-$electrons are screening themselves. 
In fact, the local ZR singlet formation does not take place 
for $\delta>0$ simply because there are no holes available for
screening. 
Nevertheless, despite the high filling of the $p-$orbitals, 
the strong hybridization $t_{pd}$ still allows for delocalization of the
$d-$electrons, through charge fluctuations across the $p-$sites, with an
effective amplitude $t_{\rm eff}\sim t_{pd}^2/\Delta$.
The key physical point in this process is that
since the $p-$orbitals are almost full, 
they have a negligible local magnetic moment, 
so these charge fluctuations take place 
without significant magnetic $p-d$ coupling. 
Therefore the magnetic phase coherence 
of the $d-$electrons is preserved and, 
in consequence, a superexchange mechanism 
between neighboring $d$ sites occurs. 
Thus, from the point of view of the $d-$electrons, 
they have strong magnetic moments, 
they delocalize keeping their quantum mechanical phase 
through essentially non-magnetic $p$-sites, 
and therefore also experience antiferromagnetic correlations 
with nearest neighbor $d-$sites. 
These physical ingredients 
are evidently also realized for the carriers 
in the single band Hubbard model. 
Therefore we can now rationalize the underlying mechanism 
for the mapping of the PAM onto 
the one band Hubbard model at $\delta> 0$, cf. Fig.~\ref{fig14a}. 

On the other hand, the situation is very different 
as the chemical potential is lowered 
to dope holes into the system. 
There, the number of available holes become more significant 
and they can lock with the robust $d-$magnetic moments 
to form local ZR singlets. 
However, when the $d-$electrons of these singlets want to delocalize, 
i.e. hop to the neighboring sites 
and eventually form a quasiparticle band, 
they completely loose the information of their spin phase, 
thus the antiferromagnetic correlations 
between neighboring $d$ do not build up. 
In consequence, Hubbard model like physics does not take place 
and the nature of both the MIT 
and the ensuing correlated metallic state 
becomes fundamentally different. 
That is the key physical reason why the mapping 
of the PAM with a one band HM is no longer valid, cf.
Fig.~\ref{fig14b}.

\begin{figure}%[!ht]
\centering
\includegraphics[width=8cm,clip]{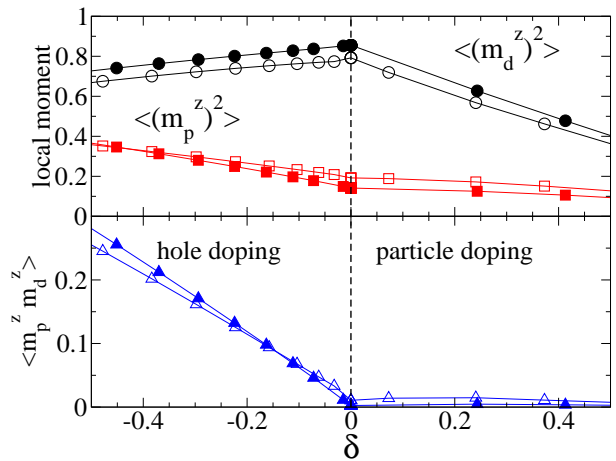}
\caption{
Top panel: $\langle (m_d^z)^2 \rangle$ (circles) 
and $\langle (m_p^z)^2 \rangle$ (squares) 
as a function of doping $\delta$ 
for $\Delta_0=1$, $t_{pd}=0.9$, 
$U=2$ (full symbols) 
and $U=1.2$ (open symbols).
Bottom panel: 
$\langle m^z_p m^z_d \rangle$
as a function of doping $\delta$,
for the same values of the parameters 
as in he main panel.
The results are obtained 
with ED with star geometry. 
}
\label{fig15}
\end{figure}

In order to substantiate the previous qualitative discussion, 
we show in Fig.~\ref{fig15} 
the $d$ and $p$ local moment formation as a function of doping 
for a large strength of the interaction $U$ 
and for a value slightly above the tip.
The local moment formation is defined as
\begin{equation}
\langle (m_{\alpha}^z)^2 \rangle =
\langle (n_{\alpha\uparrow}-n_{\alpha\downarrow})^2 \rangle = 
n_{\alpha} -2\langle n_{\alpha\uparrow} n_{\alpha\downarrow}\rangle,
\label{eq:moments}
\end{equation}
where $\alpha={p,d}$. Notice that the difference between 
the particle occupation $n$ and the moment 
directly measures the double occupancies of the sites. 
In the Mott insulator ($\delta=0$) 
the local moment of the $d$ electrons, 
$\langle (m_d^z)^2 \rangle$, is large 
because the $d-$sites are predominantly single occupied 
due to the effect of $U$. 
On the other hand, 
the $p-$band is almost fully occupied, 
and thus the local moment of the $p-$electron, 
$\langle (m_p^z)^2 \rangle$, 
is significantly smaller. 
As one dopes the Mott insulator with particles or holes, 
the relative distribution of the local moments
among both $p$ and $d$ sites is strikingly asymmetric.
Upon particle doping, $\delta>0$, 
$\langle (m_p^z)^2 \rangle$ becomes even smaller, 
since the occupation of the $p$ band get saturated. 
On the other hand, 
the $d$ moment decreases more rapidly (and linearly) 
with the doping, 
since the charge fluctuations between 
the single occupied and double occupied $d$ states 
increase. 
Therefore these results support the view that 
for the particle doped side of the transition 
the $p-$sites are magnetically inert 
and, consequently, the $d$-electrons screen 
themselves as they form a heavy mass metal. This heavy
metal paramagnetic state is physically analogous to the
one realized in the Hubbard model case. 
For the hole doped metal, $\delta<0$, 
the $p-$electron local moment, 
$\langle (m_p^z)^2\rangle$, 
increases linearly with the doping
because holes are mostly added to $p-$sites. 
In contrast, the opposite behavior occurs for the $d$ local moment, 
which slightly linearly decreases with hole doping. 
Thus, the increase of the magnetic character of $p-$electrons 
is consistent with our argument for the formation of 
local singlets in the hole doped case. 

To fully underpin our hypothesis 
we compute the magnetic moment correlation 
between the $d-$ and $p-$sites,  
\begin{equation}
\langle m^z_p m^z_d \rangle=
\langle(n_{d\uparrow}-n_{d\downarrow})
(n_{p\uparrow}-n_{p\downarrow})\rangle
\end{equation}
The results are shown in the bottom panel of Fig.~\ref{fig15}. 
They illustrate that in fact 
on the particle doped side 
the magnetic correlations are negligible, 
however on the hole doped side they increasingly grow 
as the chemical potential 
moves into the lower Hubbard band.
The growth of the expectation value 
is commensurate with the increase in hole doping 
and signals that the doped $p-$holes bind magnetically 
to the local $d-$electron magnetic moments. 
This represents the formation 
of the equivalent to Zhang Rice singlets in the present model, 
that only occurs at $\delta<0$.

\subsection{Formation of Zhang-Rice like singlets} 
The origin of the two different MIT scenarios 
and the resulting correlated metallic states 
can also be argued from an energetic point of view. 
The doping introduces new states inside the Mott gap 
$\Delta_M$, which is renormalized 
by the hybridization $t_{pd}$.
These states are a mixture of $p$ and $d$ states. 
An estimate of the energy gain 
of the Zhang Rice singlet formation is 
\begin{equation}
E_{\rm ZR} \sim \nu(1-\nu) \frac{t_{pd}^2}{\Delta \pm \Delta_M}
\label{energy_ZR}
\end{equation}
This results from the magnetic energy gain from the hopping 
of a $p-$hole ($2-n_p=\nu$) 
on a $d-$site singly occupied ($1-\nu$). 
We should compare the above binding energy 
with the delocalization energy of a $d-$electron 
between two neighboring $d-$sites, 
which is proportional to \cite{fazekas} 
\begin{equation}
E_{\rm SE} \sim (1-\nu)^2 \frac{t_{pd}^4}{\Delta^2U} = 
(1-\nu)^2 \frac{t_{\rm eff}^2}{U}
\label{energy_SE}
\end{equation}
This results from the virtual hopping of a $d-$electron
to its nearest neighbor $d$ site 
(i.e. through two $p$ sites). 
A priori this gain is of order one, 
since the $d$ site are approximatively all singly occupied, 
and $1-\nu$ electrons participate in the superexchange process
\cite{andersonSE}. 
In fact, in the region of parameter we are investigating, 
the parameters $t_{pd}$, $\Delta$ and $U$ are of order one. 
Therefore, the energy gain in the delocalization of a $d-$electron (\ref{energy_SE}) is of order $(1-\nu)^2$. 

Upon particle doping, 
the energy gain of singlet formation 
(\ref{energy_ZR}) is of order $\nu(1-\nu)$, 
thus for small but finite $\nu$, 
it is much smaller than the spin exchange energy gain (\ref{energy_SE}). 
Thus we can understand that for particle doping 
the spin exchange energy gain dominates 
on the energy gain of singlet formation. 
This imply that the metallic state 
realized upon particle doping, 
is due to the delocalization of mostly $d$ electrons. 
Therefore the $d$ electrons play the same role 
as the single type of carriers in the single band Hubbard model 
and produce qualitatively similar MIT scenario. 

On the other hand, upon hole doping, 
the energy gain of Zhang Rice singlet formation 
can be substantial. 
In fact, in our region of parameters, 
$\Delta$ is of the same order of $\Delta_M$, 
and the partial cancellation of the denominator 
in (\ref{energy_ZR}) explains 
that for a small but finite $\nu$ 
(which is controlled by the hybridization), 
the energy gain of singlet formation 
in the hole doped transition case dominates 
on the spin exchange energy gain.  
Thus we can understand that, upon hole doping, 
the ensuing metallic state is due to the delocalization 
of these composite objects. 
The nature of this metallic state 
has been studied in detail in our recent work \cite{asr08} where
we found that it severely deviated from the Fermi liquid paradigm.
The physical reason is that the delocalized holes undergo a strong magnetic
scattering from the local magnetic moments as they hop from site to site.
Thus no coherent behavior for hole propagation is realized, at least down to 
very low temperature scales, leading to the observed non-Fermi liquid character.

\subsection{Restoring the mapping of PAM onto Hubbard model}
\label{subsec:mapping}

\begin{figure}%[!ht]
\centering
\includegraphics[width=8cm,clip]{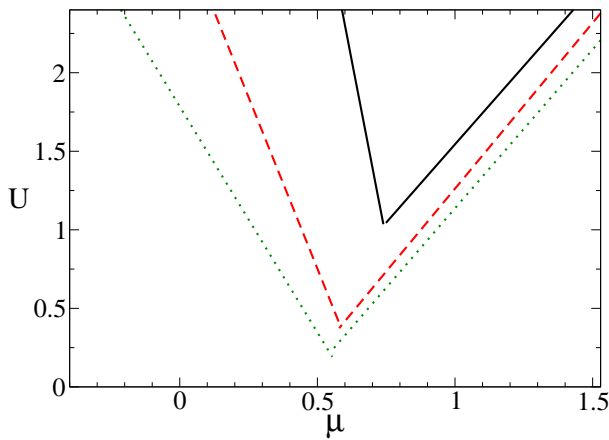}
\caption{
Phase diagram of the PAM in the $U-\mu$ plane 
that maps the Mott insulator  
and its boundaries with the 
metallic state. 
Data are for $t_{pd}=0.9$ and $\Delta_0=1,2,3$, 
i.e. $\epsilon_p=-1,-2,-3$ 
(solid, dashed and dotted lines). 
Data are obtained with $T=0$ ED calculations.
}
\label{fig16}
\end{figure}

We have shown that the new physics in the PAM 
with respect to the Hubbard model 
comes from the local coupling 
between the $p$ and $d$ electrons. 
Thus we expect that disfavoring this binding 
may restore the validity of the mapping on the hole
doped side of the MIT. 
To test this hypothesis, 
we lower the energy position of the $p-$band, 
$\epsilon_p$ 
(i.e. we increase the charge transfer energy $\Delta_0$). 
We already noticed in Sec.~\ref{subsec:mott_insulator}
that upon increasing the bare charge transfer energy 
$\Delta_0=\epsilon_d-\epsilon_p$, 
the Mott gap approaches the bare value $U$ and 
the mix-valence character of the electrons 
is in fact decreased. 

In Fig.~\ref{fig16} we show the phase diagram 
in the plane $U-\mu$ for different values 
of the position of the $p$ band. 
Upon increasing $\Delta_0$, 
the threshold value of $U$ 
to obtain the Mott insulator region 
becomes smaller. 
This results from the fact that the interaction $U$ 
is competing with a decreasing bandwidth 
$\sim t_{pd}^2/\Delta$.
In addition, as expected, 
the V-shaped boundaries of the Mott insulator 
become more symmetric 
at larger $\Delta_0$. 

To verify whether 
the character of the transition can be modified 
on the hole doped side, 
we observe the temperature behavior of 
the particle number, $n$.
Upon increasing $\Delta_0$, 
we find that $n$ versus $\mu$ curves becomes  
more temperature dependent, 
which is a first indication of the possible realization 
of the Hubbard model scenario also in the hole doped side. 
However, due to the reduction of the effective bandwidth 
$t_{eff}\sim t_{pd}^2/\Delta$, 
the temperature below which 
we may observe the hysteresis cycle 
in the $n$ versus $\mu$ curves should be extremely low. 
With our current numerical capabilities we were only 
able to obtain evidence of a small hysteresis 
at $\Delta_0=3$.

\begin{figure}%[!ht]
\centering
\includegraphics[width=8cm,clip]{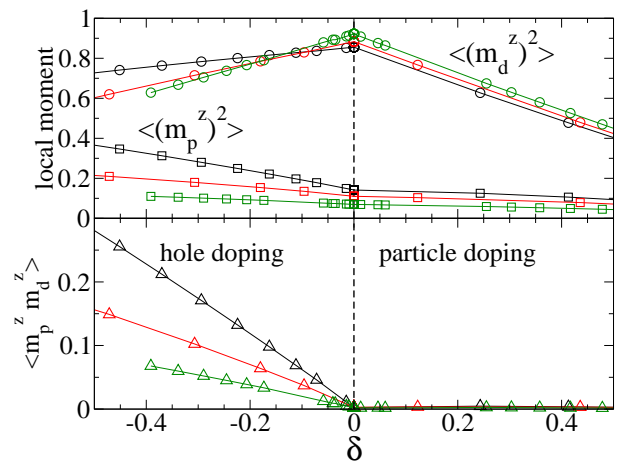}
\caption{
Top panel: 
$\langle (m_d^z)^2 \rangle$ (circles) 
and $\langle (m_p^z)^2 \rangle$ (squares) 
as a function of doping $\delta$ 
for $t_{pd}=0.9$ 
and $U$ approximately two times larger 
than the threshold value to have a Mott gap 
(tip in the V shape of the phase diagram)
and $\epsilon_p=-1,-2,-3$ 
(black, red and green symbols).
Explicitly:  
$\epsilon_p=-1$ and $U=2$,
$\epsilon_p=-2$ and $U=0.8$,
$\epsilon_p=-3$ and $U=0.4$.
Bottom panel: 
$\langle m^z_p m^z_d \rangle$
as a function of doping $\delta$,
for the same values of the parameters 
as in the top panel.
The results are obtained 
with $T=0$ ED star geometry calculations. 
}
\label{fig17}
\end{figure}

To complete this study we also computed 
the local moment of $p-$ and $d-$electrons 
upon increasing $\Delta_0$. 
In Fig.~\ref{fig17} we plot the moments (top panel) 
and the magnetic binding between the $p-$ and $d-$electrons 
(bottom panel) 
as a function of doping 
for several values of the position of the $p-$band. 
In the Mott insulator ($\delta=0$), 
upon decreasing the position of the $p$ band, 
$\langle (m_p^z)^2 \rangle$ is decreased 
and $\langle (m_d^z)^2 \rangle$ is increased. 
This is because the mix-valence character of the electrons 
is reduced, since the $p$ band become essentially full 
with $n_p\rightarrow 2$ as $\epsilon_p\rightarrow -\infty$. 
Thus the doping 
has just a small effect on 
$\langle (m_p^z)^2 \rangle$. 
On the other hand, upon hole doping,
the $d$ local moments decreases more rapidly 
when the $p-$band is deep in energy. 
As a result, the magnetic correlation 
between the $p-$ and $d-$electrons, 
$\langle m^z_p m^z_d \rangle$, 
shown in the bottom panel of Fig.~\ref{fig17},
is expectedly punished by 
higher values of $\Delta_0$ .  
This is fully consistent with the mentioned recovery 
of the mapping of the PAM onto the Hubbard model. 
Therefore we can understand that 
for lowering in energy the position of the $p-$band 
the mapping of the PAM onto the HM should hold.

\section{CONCLUSIONS}\label{sec:conclusions}

In this paper, using DMFT, we considered 
the doping-driven Mott transitions 
of the periodic Anderson model when it is 
set in the Mott-Hubbard parameter regime. We discussed the transitions 
with respect to reference case of the well understood scenario 
realized in the single band Hubbard model.

In contrast to the latter, 
the PAM has a qualitatively different 
metal-insulator transition 
for particle or hole doping. 
Upon particle doping of the Mott insulator, 
the metallic state is reached 
through a first order transition, 
that is analogous to that of the Hubbard model. 
However, upon hole doping the Mott insulator, 
there is a continuous (i.e. second order) 
insulator-metal transition through a quantum critical line 
in the parameter space $U-\mu$. 

We argued that the hole doped metal has  
delocalized Zhang Rice singlets 
that fail to build substantial superexchange 
as compared to the Hubbard model 
(and particle doped case).
In fact, we discussed the qualitative differences 
between these two transitions, 
showing that it is not due to the physics of the ``exhaustion'', 
but indeed is related to the magnetic interaction 
that develops between 
the two species of electrons in the model. 
Our results on the magnetic correlation 
between the $d-$ and the $p-$electrons 
(see lower panel of Fig.~\ref{fig15}) 
show that upon particle doping 
the $p-$electrons permit the charge fluctuations 
and the delocalization of the $d-$electrons 
without magnetic $p-d$ coupling. 
On the hole doped case, in contrast, 
the system favors the formation of singlet pairs $p-d$.

Upon increasing the charge transfer energy, 
we could recover the mapping of the PAM 
to the Hubbard model for the hole doped case. 
This signifies that a substantial mix-valence character 
was the key ingredient for the realization 
of the second order transition in this model. 

Our findings may be important 
looking at the present effort to apply DMFT calculation 
in regard to real materials \cite{phy_today,kotliarrmp}.  
Those studies usually carry the implicit assumption of 
the Hubbard model as the underlying low energy Hamiltonian 
of complex systems. 
Our work indicates 
that the Hubbard model scenario 
may be questionable when the hybridization of the correlated band with another band is high. 
In particular, our work is relevant 
for the analysis of the metal-insulator transitions 
of transition-metal oxides, 
that usually have oxygen orbital 
mediating the delocalization of the $d$ correlated electrons 
of the transition-metal.
Therefore the role of the oxygens band 
and their hybridization with the localised band 
should be explicitly considered 
in the investigation of the Mott transition.

\section*{ACKNOWLEDGMENTS}
We acknowledge M. Gabay for useful discussions.
AA acknowledges support from the European ESRT Marie-Curie program,
GS and MJR acknowledge support from the ECOS Sud-Secyt Program.

\end{document}